\documentclass[12pt, draftcls, onecolumn]{IEEEtran}

\usepackage{amsmath,amssymb,citesort}
\usepackage{graphicx}
\usepackage{subfigure}
\usepackage{color}
\usepackage{footnote}

\newtheorem{theorem}{Theorem}
\newtheorem{proposition}{Proposition}
\newtheorem{lemma}{Lemma}
\newtheorem{corollary}{Corollary}
\newtheorem{remark}{Remark}

\begin{document}

\title{Multiuser Joint Energy-Bandwidth Allocation with Energy Harvesting - Part II: Multiple Broadcast Channels \& Proportional Fairness}

\author{Zhe~Wang,
        Vaneet~Aggarwal,~
       and Xiaodong~Wang~
\thanks{The authors are with the Electrical Engineering Department, Columbia University, New York, NY 10027 (e-mail:
\{zhewang, wangx\}@ee.columbia.edu,vaneet@alumni.princeton.edu).}
}
\maketitle

\vspace{-0.39in}

\begin{abstract}
In this paper, we consider the energy-bandwidth allocation for a network with multiple broadcast channels, where the transmitters access the network orthogonally on the assigned frequency band and each transmitter communicates with multiple receivers orthogonally or non-orthogonally. We assume that the energy harvesting state and channel gain of each transmitter can be predicted for $K$ slots {\em a  priori}. To maximize the weighted throughput, we formulate an optimization problem with $O(MK)$ constraints, where $M$ is the number of the receivers, and decompose it into the energy and bandwidth allocation subproblems. In order to use the iterative algorithm proposed in \cite{OURFDPAPER} to solve the problem, we propose efficient algorithms to solve the two subproblems, so that the optimal energy-bandwidth allocation can be obtained with an overall complexity of ${\cal O}(MK^2)$, even though the problem is non-convex when the broadcast channel is non-orthogonal. For the orthogonal broadcast channel, we further formulate a proportionally-fair (PF) throughput maximization problem and derive the equivalence conditions such that the optimal solution can be obtained by solving a weighted throughput maximization problem. Further, the algorithm to obtain the proper weights is proposed. Simulation results show that the proposed algorithm can make efficient use of the harvested energy and the available bandwidth, and achieve significantly better performance than some heuristic policies for energy and bandwidth allocation. Moreover, it is seen that with energy-harvesting transmitters, non-orthogonal broadcast offers limited gain over orthogonal broadcast.
\end{abstract}


\section{Introduction}
The rapid development of energy harvesting technologies enables a new paradigm of wireless communications powered by renewable energy sources \cite{EHSNSI}\cite{Energy_Scav}.   Although energy harvesting can potentially enable sustainable and environmentally friendly deployment of wireless networks, it requires efficient utilization of energy and bandwidth resources \cite{SROPPE}\cite{ALLERTON}. 

In Part I of this two-part paper \cite{OURFDPAPER}, for a network with multiple orthogonal broadcast channels and energy harvesting transmitters, we proposed an iterative algorithm for computing the optimal energy-bandwidth allocation to maximize the weighted throughput. For the special case that each transmitter only communicates with one receiver and all weights are equal, the algorithms for efficiently solving the energy and bandwidth allocation subproblems are also proposed. In this paper, we develop algorithms for solving the two subproblems for the general case of multiple broadcast channels. Moreover, for a single (non-orthogonal) broadcast channel with energy harvesting transmitter, the optimal energy scheduling over static and two-user fading channels was discussed in \cite{BEHRT}  and \cite{OSFBC}, respectively. In this paper, we treat the energy-bandwidth allocation problem for multiple broadcast channels, including both orthogonal and non-orthogonal broadcast. Taking the proportional fairness into account, \cite{CPFSAG} discussed the convergence of the general proportionally-fair scheduling without energy harvesting. For energy harvesting transmitters with unbounded battery capacity, heuristic algorithms have been proposed  in \cite{PFRAEH} to find the time-power allocations under the proportional fairness. The proportionally-fair energy-bandwidth allocation in multiple orthogonal broadcast channels is also treated in this paper. 

In particular, we consider a network with multiple transmitters, each powered by the renewable energy source. We assume that the transmitters are assigned orthogonal frequency bands to avoid interfering from each other. In orthogonal broadcast, the frequency band assigned to the transmitter is further split for the transmission to each designated receiver orthogonally (i.e., no interference); on the other hand, in non-orthogonal broadcast, the transmissions to all designated receivers take place on the same frequency band assigned to the transmitter. For the special case where all links have equal weights, with orthogonal or non-orthogonal broadcast, we show that each transmitter should only use the strongest channel in each slot, i.e., multiple broadcast channels reduce to multiple point-to-point channels, and thus we can  directly use the algorithms in \cite{OURFDPAPER} to obtain the optimal energy-bandwidth allocation. For the general weighted case, we develop algorithms for solving the two subproblems, i.e., energy allocation and bandwdith allocation, for both orthogonal and non-orthogonal broadcast. We also reveal that the gain by non-orthognoal broadcast over orthogonal broadcast is limited with energy harvesting transmitters.

Moreover, we formulate a proportionally-fair (PF) throughput maximization problem with orthogonal broadcast. In point-to-point channels without energy harvesting, in slot $k$, the optimal PF scheduler schedules the link with $\max_m R_m^k/A_m^k$, where $R_m^k$ is the rate achievable by link $m$ in slot $k$ and $A_m^k$ is the average rate of link $m$ up to slot $k$. The average rate is computed over a time window as a moving average: $R_m^{k+1} = (1-\alpha)A_m^k+\alpha R_m^k$ if link $m$ is scheduled in slot $k$, and $A_m^{k+1} = (1-\alpha)A_m^k$ otherwise \cite{CPFSAG}. However, in the presence of energy harvesting, using a single link is not optimal and thus scheduling multiple links in a slot and splitting the bandwidth is essential. To efficiently solve the PF throughput maximization problem, we convert it to a weighted throughput maximization problem with proper weights. The algorithm to obtain such weights is also proposed.

The remainder of the paper is organized as follows. Sections II and III treat orthogonal and non-orthogonal broadcast channels, respectively. Section IV solves the propotionally fair problem for orthogonal broadcast. Simulation results are provided in Section V. Finally, Section VI concludes the paper.

\section{Multiple Orthogonal Broadcast Channels}
Consider a network consisting of $N$ transmitters and $M$ receivers where transmitter $n\in{\cal N}$ communicates with receivers in the set ${\cal M}_n$ ($\bigcup_n {\cal M}_n = {\cal M}$, and ${\cal M}_n\bigcap {\cal M}_{n'}=\Phi$ for $n\neq n'$) in an orthogonal broadcast channel. Our goal is to schedule the transmission in $K$ slots  ${\cal K}\triangleq \{1,2,\ldots, K\}$  to maximize the weighted sum-rate by proper energy and bandwidth allocation \cite[Eqn. (5)-(6)]{OURFDPAPER}. Specifically, in \cite{OURFDPAPER}, we first gave the optimal energy discharge schedule in \cite[Eqn. (11)]{OURFDPAPER} and then proposed an iterative algorithm \cite[Algorithm 1]{OURFDPAPER} to obtain the optimal energy allocation ${\cal P}\triangleq \{p_{m}^k,\forall m\in{\cal M},k\in{\cal K}\}$ and the bandwidth allocation ${\cal A}\triangleq\{a_{m}^k,\forall m\in{\cal M},k\in{\cal K}\}$.

Recall the general energy-bandwidth allocation problem  ${\sf P}_{\cal W}(\epsilon)$ for multiple orthogonal broadcast channels formulated in \cite[Eqn. (12)-(13)]{OURFDPAPER}: 
\begin{equation}\label{eq:weighedproblem}
{\sf P}_{\cal W}(\epsilon):\quad \max_{{\cal P},{\cal A}} C_{\cal W}({\cal P,A})
\end{equation}
subject to
\begin{equation}\label{eq:cst}
\left\{
\begin{array}{l}
\tilde{E}_n^k - B_n^{\max} \leq  \sum_{\kappa=1}^{k}\sum_{m\in{\cal M}_n}p_m^{\kappa} \leq \tilde{E}_n^k \\
\sum_{m=1}^M a_m^k = 1\\
\sum_{m\in{\cal M}_n} p_m^k \leq P_n\\
p_m^k \geq 0\\
a_m^k \geq \epsilon
\end{array}\right.
\end{equation}
for all $n\in{\cal N}, m\in{\cal M}, k\in{\cal K}$, where 
\begin{equation}
C_{\cal W}({\cal P},{\cal A}) = \sum_{m\in{\cal M}}W_{m}\sum_{k\in{\cal K}} a_m^k \log(1+\frac{p_m^kH_m^k}{a_m^k}),\ a_m^k\in[0,1], p_m^k\in[0,\infty),
\end{equation}
${\cal W}\triangleq\{W_{m},\forall m\in{\cal M}\}$ is the set of weights, $\epsilon$ is  the required minimal bandwidth allocation, $\tilde{E}_n^k$ is the effective harvested energy  after optimally discharging the surplus energy \cite[Eqn. (11)]{OURFDPAPER}, and $B_n^{\max}$ is the battery capacity of transmitter $n$.

Introducing the non-negative dual variables $\lambda_n^k$, $\mu_n^k$, $\alpha^k$, $\beta_m^k$ and $\xi_n^k$ for all $n\in{\cal N}, m\in{\cal M}$ and $k\in{\cal K}$, we denote 
\begin{align}\label{eq:multiplier}
{\cal M}({\cal P,A}) \triangleq& -\sum_{n,k}\lambda_n^k \left( \sum_{\kappa=1}^{k}\sum_{m\in{\cal M}_n}p_m^{\kappa} - \tilde{E}_{n}^k \right)+\sum_{n,k}\mu_n^k \left( \sum_{\kappa=1}^{k}\sum_{m\in{\cal M}_n}p_m^{\kappa} - \tilde{E}_n^k + B_n^{\max} \right)\nonumber\\
&-\sum_{k}\alpha^k(\sum_{m}a_{m}^k-1)+\sum_{m,k}\beta_{m}^k(a_{m}^k-\epsilon) - \sum_{n,k} \xi_n^k( \sum_{m\in{\cal M}_n}p_{m}^k - P_n)\nonumber\\
=&-\sum_{n,k}\left(\sum_{m\in{\cal M}_n}p_{m}^{k} \sum_{\kappa=k}^K \lambda_{n}^{\kappa} - \lambda_{n}^{k}\tilde{E}_n^{k} \right)+\sum_{n,k}\left(\sum_{m\in{\cal M}_n}p_{m}^{k} \sum_{\kappa=k}^K \mu_{n}^{\kappa} -\mu_n^k\left(\tilde{E}_n^{k}-B_n^{\max}\right) \right)\nonumber\\
&-\sum_{k}\alpha^k(\sum_{m}a_{m}^k-1)+\sum_{m,k}\beta_{m}^k(a_{m}^k-\epsilon) - \sum_{n,k} \xi_n^k( \sum_{m\in{\cal M}_n}p_{m}^k - P_n)\ ,
\end{align}
as the Lagrangian multipliers. Then, the Lagrangian functions for ${\sf P}_{\cal W}(\epsilon)$ can be defined as 
\begin{equation}\label{L:O}
{\cal L}_O\triangleq C_{\cal W}({\cal P,A})+ {\cal M}({\cal P,A})\ .
\end{equation}

\subsection{Maximizing Network Throughput}
For the special case that all links have equal weights, e.g., ${\cal W}=\{W_m=1,m\in{\cal M}\}$, the following result states that each transmitter should only use its strongest channel.

\begin{theorem}\label{thm:ew}
The problem $P_{\{1\}}(0)$ in multiple orthogonal broadcast channels is equivalent to the energy-bandwidth allocation problem in point-to-point channels formulated as
\begin{equation}
\max_{{\cal P},{\cal A}} \sum_{n\in{\cal N},k\in{\cal K}} a_{m_n^k}^k\log\left(1 + \frac{p_{m_n^k}^k H_{m_n^k}^k}{ a_{m_n^k}^k}\right)
\end{equation}
subject to the constraints in \eqref{eq:cst}, where $m_n^k\triangleq\arg\max_{m\in{\cal M}_n}\{H_m^k\}$ for each $k\in{\cal K}$. Thus the optimal energy-bandwidth allocation can be efficiently solved  by the algorithms in \cite{OURFDPAPER}.
\end{theorem}

\begin{IEEEproof}
The first-order condition is necessary for optimality, which can be written as
\begin{align}
\frac{H_m^k}{1+p_m^kH_m^k/a_{m}^k} &= \frac{v_n^k - u_n^k + \xi_n^k}{W_m},\ m\in{\cal M}_n\label{kkt:wf}\ ,\\
\textrm{with }\quad\quad\quad\quad u_n^k &\triangleq \sum_{\kappa=k}^K \mu_n^{\kappa}\ ,\nonumber\\
v_n^k &\triangleq \sum_{\kappa=k}^K \lambda_n^{\kappa}. \label{eq:level}
\end{align}

By setting $W_m=1$, we then have
\begin{equation}\label{eq:ewea}
p_m^k = a_{m}^k  \left[\frac{1}{v_n^k - u_n^k + \xi_n^k} - \frac{1}{H_m^k}\right]^+\ .
\end{equation}

When $\sum_{m\in{\cal M}} p_m^k > 0$ and $\epsilon=0$,  the optimal bandwidth allocation is given as \cite{OEBAEH}
\begin{equation}
a_{m}^k = \frac{p_m^kH_m^k}{\sum_{j\in{\cal M}} p_{j}^kH_{j}^k}, \ m\in{\cal M}\ .
\end{equation}
Then, for any transmitter $n$ such that $\sum_{m\in{\cal M}_n} p_m^k > 0$ and denoting $\Delta \triangleq \sum_{m\in{\cal M}_n} a_m^k$, we further have
\begin{equation}\label{eq:ewba}
a_{m}^k =  \frac{p_m^kH_m^k \Delta}{\sum_{j\in{\cal M}_n} p_{j}^kH_{j}^k}, \ m\in{\cal M}_n\subseteq{\cal M} \ .
\end{equation}
Substituting \eqref{eq:ewba} into \eqref{eq:ewea}, we then have
\begin{align}
p_m^k &= \frac{p_m^kH_m^k}{\sum_{j\in{\cal M}} p_{j}^kH_{j}^k}  \left[\frac{1}{v_n^k - u_n^k + \xi_n^k} - \frac{1}{H_m^k}\right]^+ \Delta,\ m\in{\cal M}_n\ . \label{eq:ewtemp}
\end{align}
Replacing $p_j^k$ in \eqref{eq:ewtemp} by \eqref{eq:ewea}, we have
\begin{align}
p_m^k &= p_m^k \frac{ \left[\frac{1}{v_n^k - u_n^k + \xi_n^k} - \frac{1}{H_m^k}\right]^+ H_m^k\Delta}{a_m^k  \left[\frac{1}{v_n^k - u_n^k + \xi_n^k} - \frac{1}{H_m^k}\right]^+ H_m^k + \sum_{j\in{\cal M}_n,j\neq m}  a_j^k\left[\frac{1}{v_n^k - u_n^k + \xi_n^k} - \frac{1}{H_{j}^k}\right]^+ H_{j}^k} \ .\label{eq:temp}
\end{align}
When $p_m^k > 0$,  $\left[\frac{1}{v_n^k - u_n^k + \xi_n^k} - \frac{1}{H_m^k}\right]^+ > 0$ and \eqref{eq:temp} can be further written as
\begin{align}
1&=\frac{\Delta}{a_m^k+\left( \sum_{j\in{\cal M}_n,j\neq m}a_j^k   \left[\frac{1}{v_n^k - u_n^k + \xi_n^k} - \frac{1}{H_{j}^k}\right]^+ H_{j}^k\right) /\left ( \left[\frac{1}{v_n^k - u_n^k + \xi_n^k}- \frac{1}{H_m^k}\right]^+ H_m^k\right)}\ ,\\
\Rightarrow \quad \Delta &=a_m^k+\sum_{j\in{\cal M}_n,j\neq m}a_j^k\left( \frac{ \left[1/(v_n^k - u_n^k + \xi_n^k) - 1/H_{j}^k\right]^+}{\left[1/(v_n^k - u_n^k + \xi_n^k)- 1/H_m^k\right]^+}\frac{H_{j}^k}{ H_m^k}\right)\ .\label{eq:eqg1}
\end{align}
Moreover, according to the definition of $\Delta$, we also have
\begin{equation}\label{eq:eqg2}
 a_m^k+\sum_{j\in{\cal M}_n,j\neq m}a_j^k \cdot 1 = \Delta\ .
\end{equation}

Denoting $m_n^k\triangleq \max_{m\in{\cal M}_n} \left\{H_m^k\right\}$, by \eqref{eq:ewea} and \eqref{eq:ewba}, we have 
$p_{m_n^k}^k > 0$ when $\sum_{m\in{\cal M}_n}p_m^k > 0$. Note that, since
\begin{equation}
 \frac{ \left[1/(v_n^k - u_n^k + \xi_n^k) - 1/H_{j}^k\right]^+}{\left[1/(v_n^k - u_n^k + \xi_n^k)- 1/H_{m_n^k}^k\right]^+}\cdot\frac{H_{j}^k}{ H_{m_n^k}^k} \leq 1
\end{equation}
for all $j\in\{m\in{\cal M}_n\;|\;m\neq m_n^k\}$, we must have $a_j^k=0$ for all $j\in\{m\in{\cal M}_n\;|\;m\neq m_n^k\}$ so that \eqref{eq:eqg1} and \eqref{eq:eqg2} are both satisfied. 

Therefore, when $\sum_{m\in{\cal M}_n} p_m^k > 0$, we must have $p_{m_n^k}^k >0$ and $p_{j}^k=0$ for $\{\forall j\in{\cal M}_n\;|\;j\neq m_n^k\}$. On the other hand, when $\sum_{m\in{\cal M}_n} p_m^k = 0$, we have $p_m^k=0$ for all $m\in{\cal M}_n$ given $n$ and $k$ thus the achievable rate is zero no matter which channel is selected. 
\end{IEEEproof}


\subsection{Optimal Algorithms for Solving Subproblems}
For the general weighted sum-rate problem, the iterative algorithm developed in \cite{OURFDPAPER} decomposes ${\sf P}_{\cal W}(\epsilon)$ as follows. 
\begin{itemize}
\item Given the bandwidth allocation  ${\cal A}_n\triangleq \{a_{m}^k,\forall m\in{\cal M}_n,k\in{\cal K}\}$, for each $n \in{\cal N}$, obtain the energy allocation $\boldsymbol{p}_m\triangleq[p_m^1,p_m^2,\ldots,p_m^K]$ by solving the following subproblem:
\begin{equation}
{\sf EP}_{n}({\cal A}_n,{\cal W}):\quad \max_{\boldsymbol{p}_m,m\in{\cal M}_n} \sum_{m\in{\cal M}_n}W_m \sum_{k=1}^K  a_m^k \log(1+\frac{p_m^kH_m^k}{a_m^k})
\end{equation}
subject to
\begin{equation}
\left\{
\begin{array}{l}
\tilde{E}_n^k - B_n^{\max} \leq  \sum_{\kappa=1}^{k}\sum_{m\in{\cal M}_n}p_m^{\kappa} \leq \tilde{E}_n^k \\
\sum_{m\in{\cal M}_n} p_m^k\leq P_n\\
p_m^k \geq 0,\ m\in{\cal M}_n\\
\end{array}\right. ,\ k\in{\cal K}\ .
\end{equation}
\item Given the energy allocation ${\cal P}_k\triangleq \{p_m^k,\forall m\in{\cal M}\}$, for each $k \in{\cal K}$, obtain the bandwidth allocation $ \boldsymbol{a}^k\triangleq[a_1^k,a_2^k,\ldots,a_M^k]$ by solving the following subproblem:
\begin{equation}
{\sf BP}_k({\cal P}_k,\epsilon,{\cal W}):\quad\max_{\boldsymbol{a}^k}\sum_{m=1}^M W_m a_m^k \log(1+\frac{p_m^kH_m^k}{a_m^k})
\end{equation}
subject to
\begin{equation}\label{eq:bpcst}
\left\{
\begin{array}{l}
\sum_{i=1}^M a_i^k = 1\\
a_m^k \geq \epsilon,\ m\in{\cal M}
\end{array}\right. \ .
\end{equation}
\end{itemize}

In \cite{OURFDPAPER}, algorithms for solving the above two subproblems are obtained for the special case of point-to-point channels and equal weights. We now develop algorithms for the general case.



\subsubsection{Solving the Bandwidth Allocation Subproblem}

Based on the Lagrangian function defined in \eqref{L:O}, the first-order condition and the complementary slackness of the bandwidth allocation problem can be written as
\begin{align}
&\log(1+\frac{p_{m}^kH_{m}^k}{a_{m}^k}) - \frac{p_{m}^kH_{m}^k}{a_{m}^k + p_{m}^kH_{m}^k} = \frac{(\alpha^k-\beta_{m}^k)}{W_m}\label{kkt:al}\ ,\\
&\alpha^k (\sum_{m} a_{m}^k - 1) = 0,\label{kkt:c}\\
&\beta_{m}^k ( a_{m}^k - \epsilon) = 0\label{kkt:d},
\end{align}
which along with the constraints in \eqref{eq:bpcst} constitute the K.K.T. conditions of ${\sf BP}_k({\cal P}_k,\epsilon,{\cal W})$. Since ${\sf BP}_k({\cal P}_k,\epsilon,{\cal W})$ is a convex optimization problem with linear constraints, its K.K.T. conditions are sufficient and necessary for optimality when $\epsilon >0$ \cite{CO}. 

Denote $x_{m}^k=X_{m}(\alpha^k,\beta_{m})$ as the solution to
\begin{equation}\label{eq:equation}
x_{m}^k-\log(x_{m}^k) =(\alpha^k-\beta_{m}^k)/W_m + 1\ , 0<x_{m}< 1 .
\end{equation}
Note that, for $x\in(0,1)$, $x - \log(x)\in(1,\infty)$. Then, $x_{m}^k\in(0,1)$ exists when  $\alpha^k-\beta_{m}^k \geq 0$ and the bandwidth allocation given by
\begin{equation}\label{eq:aln}
a_{m}^k = p_{m}^kH_{m}^k\frac{X_{m}(\alpha^k,\beta_{m}^k)}{1-X_{m}(\alpha^k,\beta_{m}^k)},\ (0 <  X_m(\alpha^k,\beta_{m}^k) < 1)
\end{equation}
for $p_{m}^k > 0$ satisfies the first-order condition in \eqref{kkt:al}. 


When $p_{m}^k=0$,  we have $\alpha^k = \beta_{m}^k \geq 0$ by \eqref{kkt:al}. If $\alpha^k  = \beta_{m}^k > 0$,  we have $a_{m}^k=\epsilon$ by \eqref{kkt:d}. Otherwise, we can set $a_{m}^k=\epsilon$ and the K.K.T. conditions still hold. Thus the minimal bandwidth should be assigned to the receiver with zero transmission energy.

We note that, if there exists an $m$ such that $p_{m}^k > 0$, the left-hand-side of \eqref{kkt:al} is greater than $0$ and thus $\alpha^k > 0$. Then, by \eqref{kkt:c}, $\sum_{m}a_{m}^k=1$ must hold. Assigning the minimal bandwidth to the receiver with zero transmission energy and substituting \eqref{eq:aln}, we further have
\begin{equation}\label{eq:sum1}
\sum_{m\in{\cal Z}_0^c}  p_{m}^kH_{m}^k\frac{X_{m}(\alpha^k,\beta_{m})}{1-X_{m}(\alpha,\beta_{m})} + |{\cal Z}_0| \epsilon= 1\ ,
\end{equation}
where ${\cal Z}_0\triangleq \{m\;|\;p_{m}^k=0\} = \{m\;|\; p_{m}^k=0, a_{m}^k = \epsilon\}$ and ${\cal Z}_0^c$  is the complementary set of ${\cal Z}_0$. Moreover, by \eqref{kkt:d}, we know that $\beta_{m}^k=0$ when $a_{m}^k > \epsilon$. Then, \eqref{eq:sum1} can be further written as
\begin{equation}\label{eq:optba}
\sum_{m\in{\cal Z}_1^c\cap {\cal Z}_0^c}  p_{m}^kH_{m}^k\frac{X_m(\alpha^k,0)}{1-X_m(\alpha^k,0)}  +  |{\cal Z}_1| \epsilon= 1 - |{\cal Z}_0| \epsilon\ ,
\end{equation}
where ${\cal Z}_1\triangleq \{m\;|\; p_{m}^k>0, \beta_{m}^k > 0\}$.

Note that, for any $m\in{\cal Z}_1$, we have
\begin{equation}
a_{m}^k= p_{m}^kH_{m}^k \frac{X_m(\alpha^k,\beta_{m}^k)}{1-X_m(\alpha^k,\beta_{m}^k)} = \epsilon,\ (\beta_{m}^k > 0) \ .
\end{equation}
According to \eqref{eq:equation}, since $X_m(\alpha, \beta)$ is decreasing with respect to $\alpha$ and increasing with respect to $\beta \geq 0$ when $X_m(\alpha,0)\in(0,1)$, then so does $ \frac{X_m(\alpha,\beta)}{1-X_m(\alpha,\beta)}$. Hence, we further have
\begin{equation}\label{eq:baieq}
p_{m}^kH_{m}^k \frac{X_m(\alpha^k,0)}{1-X_m(\alpha^k,0)} \leq p_{m}^kH_{m}^k \frac{X_m(\alpha^k,\beta_{m}^k)}{1-X^k_{m}(\alpha^k,\beta_{m}^k)} = \epsilon,\ m\in{\cal Z}_1 \ .
\end{equation}
Therefore, \eqref{eq:optba} can be written as
\begin{equation}\label{eq:optban}
\sum_{m\in{\cal Z}_0^c}  \max\left\{\epsilon, p_{m}^kH_{m}^k\frac{X_m(\alpha^k,0)}{1-X_m(\alpha^k,0)}\right\} = 1 - |{\cal Z}_0| \epsilon\ .
\end{equation}

\begin{theorem}\label{thm:oba}
Suppose that $\alpha^k$ is the solution to \eqref{eq:optban}. Then, the optimal bandwidth allocation for  ${\sf BP}_k({\cal P}_k,\epsilon,{\cal W})$ is given by
\begin{equation}\label{eq:oba}
a_{m}^k=\left\{\begin{array}{ll}
 \epsilon,& \textrm{ if } p_{m}^k=0\\
 \max\left\{\epsilon, p_{m}^kH_{m}^k\frac{X_m(\alpha^k,0)}{1-X_m(\alpha^k,0)}\right\},  &\textrm{ if }  p_{m}^k>0\\
\end{array}\right.\ .
\end{equation}
\end{theorem}
\begin{IEEEproof}
The first term in \eqref{eq:oba} follows since the minimal bandwidth should be allocated to the receiver with zero transmission energy. Also, by  \eqref{eq:baieq} and \eqref{eq:aln} we have the second term in \eqref{eq:oba}. Moreover, when $\alpha^k$ satisfies  \eqref{eq:optban}, all K.K.T. conditions of the bandwidth allocation problem are satisfied therefore the optimal bandwidth allocation is obtained.
\end{IEEEproof}

Denote
\begin{equation}
G(\alpha) \triangleq \sum_{m\in{\cal Z}_0^c}  \max\left\{\epsilon, p_{m}^kH_{m}^k\frac{X_m(\alpha^k,0)}{1-X_m(\alpha^k,0)}\right\}\ .
\end{equation}
We note that $X_m(\alpha^k,0)\in(0,1)$ is continuous and decreasing with respect to $\alpha^k$, then so does $\frac{X_m(\alpha^k,0)}{1-X_m(\alpha^k,0)}$. Since $p_{m}^kH_{m}^k$ is constant, we have that $G(\alpha^k)\in(0,+\infty)$ is also continuous and decreasing with respect to $\alpha^k$. Then, we may use the bisection method \cite{NA} to find out $\alpha^k$ such that $G(\alpha^k) =  1 - |{\cal Z}_0| \epsilon$ and the optimal bandwidth allocation can be obtained by \eqref{eq:oba}.

The procedure for solving the bandwidth allocation is summarized as follows.\\
\begin{minipage}[h]{6.5 in}
\rule{\linewidth}{0.3mm}\vspace{-.05in}
{\bf {\footnotesize Algorithm 1 - Solving bandwidth allocation subproblem  ${\sf BP}_k({\cal P}_k,\epsilon,{\cal W})$}}\vspace{-.1in}\\
\rule{\linewidth}{0.2mm}
{ {\small
\begin{tabular}{ll}
	\;1:&  Initialization\\
		\;& Specify initial $\alpha_u>\alpha_l> 0$ ($G(\alpha_u) <1 - |{\cal Z}_0| \epsilon<G(\alpha_l) $) and error tolerance $\delta>0$\\
    \;2:& {\bf REPEAT}\\
    \;&\quad $\alpha \leftarrow (\alpha_u+\alpha_l )/2$\\
    \;& \quad {\bf FOR} all $m\in{\cal M}$\\
      \;&\quad  \quad Calculate $X_m(\alpha,0)$ by solving \eqref{eq:equation} with $\beta=0$\\
      	\;&\quad {\bf ENDFOR}\\
      	      	\;& \quad  Evaluate $G(\alpha)$ using $\{X_m(\alpha,0),m\in{\cal M}\}$\\
      	\;&\quad {\bf IF} $|G(\alpha) - 1 + |{\cal Z}_0|| < \delta$ {\bf THEN} Goto step 3 {\bf ENDIF}\\
      	\;& \quad {\bf IF} $G(\alpha) >1 - |{\cal Z}_0| \epsilon$ {\bf THEN} $\alpha_l\leftarrow \alpha$ {\bf ELSE}  $\alpha_h\leftarrow \alpha$ {\bf ENDIF} \\
      	\;3:& {\bf FOR} all $m\in{\cal M}$\\
      	    \;&  \quad Calculate $a_{m}^k$ by \eqref{eq:oba} \\
      	\;& {\bf ENDFOR}\\
     	    
\end{tabular}}}\\
\rule{\linewidth}{0.3mm}
\end{minipage}\vspace{.05 in}

Since we need to solve  for $X_m(\alpha,0)$ from \eqref{eq:equation} repeatedly, we can pre-compute the solutions to $y=x-\log(x), \ x\in(0,1)$  and store them in a look-up table. Then the overall complexity of Algorithm 1 is ${\cal O}(M)$ for solving ${\sf BP}_k({\cal P}_k,\epsilon,{\cal W})$.

\begin{remark}
In \cite{OURFDPAPER}, we focused on the special case of equal weights, where the optimal bandwidth allocation can be directly obtained by the iterative bandwidth fitting algorithm \cite[Algorithm 2]{OURFDPAPER} without solving the dual variable $\alpha^k$ and calculating the intermediate variable $X_m(\alpha^k,0)$. However, for the general weighted case, we need to solve the equation group consisting of \eqref{eq:equation} for all $m\in{\cal M}$ and \eqref{eq:optban} to obtain the dual variable $\alpha^k$ and then calculate the optimal bandwidth allocation given by \eqref{eq:oba}.
\end{remark}

\subsubsection{Solving the Energy Allocation Subproblem}
${\sf EP}_{n}({\cal A}_n,{\cal W})$ is a convex optimization problem with linear constraints thus its K.K.T. conditions are necessary and sufficient for optimality \cite{CO}. Using the Lagrangian function defined in \eqref{L:O}, in addition to the first-order condition and the feasibility constraints,  the complementary slackness can be written as
\begin{align}
\lambda^k(\sum_{\kappa=1}^k\sum_{m\in{\cal M}_n}p_{m}^{\kappa}-{E}^k)&=0,\label{kkt:a}\\
\mu_{n}^k (\sum_{\kappa=1}^k\sum_{m\in{\cal M}_n}p_{m}^{\kappa}-{E}^k+B^{\max})&=0,\label{kkt:b}\\
\xi_n^k (\sum_{m\in{\cal M}_n}p_m^k - P_n) &= 0\label{kkt:e}
\end{align}
constituting the K.K.T. conditions. 

Taking the derivative of \eqref{eq:multiplier} on $p_m^k$ and using the first-order condition, we have
\begin{equation}\label{eq:nwf}
p_{m}^k=a_{m}^k \left[\frac{W_{m}}{v_n^k-u_n^k + \xi_n^k} - \frac{1}{H_{m}^k}\right]^+\ .
\end{equation}
By \eqref{kkt:e}, when $\sum_{m\in{\cal M}_n}p_m^k = P_n$,  we have $\xi_n^k \geq 0$ and otherwise  $\xi_n^k=0$. Then, we have
\begin{equation}
p_m^k=a_{m}^k \left[\frac{W_{m}}{v_n^k-u_n^k} - \frac{1}{H_{m}^k}\right]^+
\end{equation}
when $\sum_{m\in{\cal M}_n}a_{m}^k\left[\frac{W_{m}}{v_n^k-u_n^k} - \frac{1}{H_{m}^k}\right]^+ < P_n$. Otherwise, since the constraint requires $\sum_{m\in{\cal M}_n}p_m^k \leq P_n$, given $v_n^k$ and $u_n^k$, we can determine $\bar{\xi}_n^k\geq 0$ such that
\begin{equation}
\sum_{m\in{\cal M}_n}a_{m}^k\left[\frac{W_{m}}{v_n^k-u_n^k} - \frac{1}{H_{m}^k}\right]^+ \geq \sum_{m\in{\cal M}_n}a_{m}^k\left[\frac{W_{m}}{v_n^k-u_n^k+\bar{\xi}_n^k} - \frac{1}{H_{m}^k}\right]^+ = P_n\ .
\end{equation}
Then we can treat
\begin{equation}
\bar{P}_m^k \triangleq a_m^k \left[\frac{W_{m}}{v_n^k-u_n^k + \bar{\xi}_n^k}- \frac{1}{H_m^k}\right]^+
\end{equation}
as the maximum transmission energy for each receiver and thus the optimal energy allocation is
\begin{equation}\label{eq:nwf}
p_{m}^k=\min\left\{\bar{P}_m^k,a_{m}^k \left[W_{m} w_n^k - \frac{1}{H_{m}^k}\right]^+\right\}\ ,
\end{equation}
where $w_n^k \triangleq 1/(v_n^k-u_n^k)$.

We note that, $p_m^k$ in \eqref{eq:nwf} is a function of $w_n^k$. Then, using the same analysis in \cite{OURFDPAPER}, we have the following proposition:
\begin{proposition}\label{pp:wf}
Given any bandwidth allocation ${\cal A}_n$, $p_m^k$ is the optimal energy allocation for ${\sf EP}_n({\cal A}_n,{\cal W})$, if and only if, the feasible allocation $p_m^k$ follows the generalized two-dimensional water-filling formula in \eqref{eq:nwf}, where the water level $w_n^k$ may only increase at BDP such that $B_n^k=0$ and only decrease  at BFP such that $B_n^k=B_n^{\max}$.
\end{proposition}

\begin{figure}
\centering
\includegraphics[width=.95\textwidth]{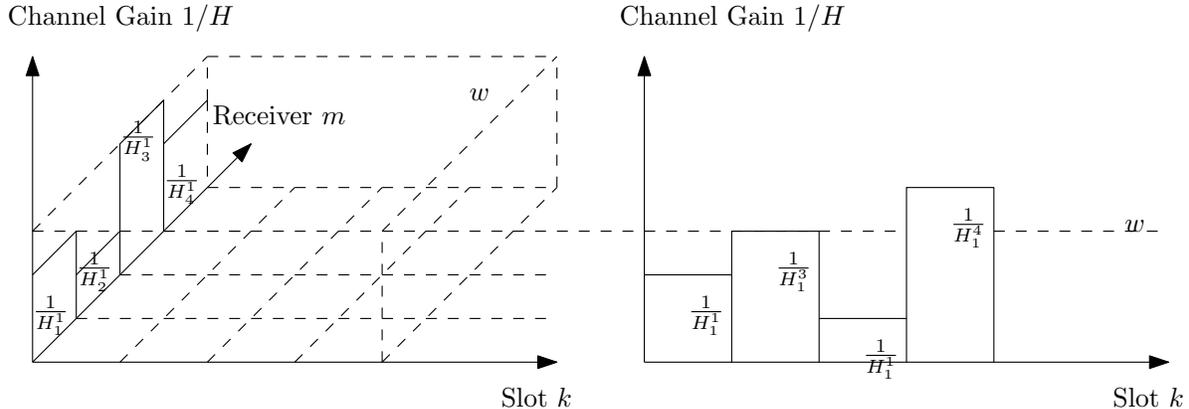}
\caption{Two-dimensional water-filling. The ``water'' (energy) is filled over both the receiver-axis (left) and time-axis (right) with the same water level $w$ as interpreted in \eqref{eq:nwf}-\eqref{eq:wl}.}
\label{fg:2d}
\end{figure}

We note that, in the orthogonal broadcast channel, each transmitter communicates with multiple receivers and the transmitted energy is drawn from the same battery. Then, according to \eqref{eq:nwf}, the water (energy) is not only filled along the time axis but also along the receiver index axis, as shown in Fig. \ref{fg:2d}. In other words,  given two  adjacent BDP/BFPs $(a,\textrm{ the type of } a)$ and $(b,\textrm{ the type of } b)$ where $a \leq b$, the energy allocation $p_m^k$ can be calculated by \eqref{eq:nwf} with the same water level $w_n^k=w^{ab}$ for all receiver $m\in{\cal M}_n$ and slot $k\in[a+1,b]$. Then, the water level $w^{ab}$ should be determined by
\begin{equation}\label{eq:wl}
\sum_{k=a+1}^b \sum_{m=1}^M p_m^k(w^{ab}) = E^b - E^a + \left(\mathbb{I}(a \textrm{ is BFP}) - \mathbb{I}(a \textrm{ is BDP}) \right) B_n^{\max}\end{equation}
where $\mathbb{I}(\cdot)$ is an indicator function and $p_m^k(w^{ab})$ is calculated by \eqref{eq:nwf} with $w_n^k = w^{ab}$ for $k\in[a+1,b]$.

In  \cite{2014arXiv1401.2376W}, a single-user dynamic water-filling algorithm is proposed to find the BDP/BFP set by recursively performing the ``forward search'' and ``backward search'' operations with conventional water-filling. Since here the increase/decrease of the water level also occurs at BDP/BFPs, replacing the conventional water-filling used in  \cite{2014arXiv1401.2376W}  by the two-dimensional water-filling in \eqref{eq:nwf}-\eqref{eq:wl},  we can obtain the BDP/BFP set for optimal energy allocation in multiple orthogonal broadcast channels. We name this algorithm as the {\em two-dimensional dynamic water-filling algorithm}. Moreover, after obtaining the optimal BDP/BFP set, the optimal energy allocation can be further calculated by \eqref{eq:nwf}-\eqref{eq:wl}.

\begin{remark}
We note that, with equal weights, by Theorem \ref{thm:ew}, the energy-bandwidth allocation problem for multiple orthogonal broadcast channels is equivalent to that for multiple point-to-point channels treated in \cite{OURFDPAPER}. Although the general algorithms developed in this section can obtain the optimal energy-bandwidth allocation for the equal weight case, solving the problem by using Theorem \ref{thm:ew} along with the algorithms in \cite{OURFDPAPER} has a lower computational complexity. Specifically, for the general case, the energy allocation subproblem ${\sf EP}_{n}({\cal A}_n,{\cal W})$ contains ${\cal O}(|{\cal M}_n|K)$ variables and the bandwidth allocation subproblem ${\sf BP}_k({\cal P}_k,\epsilon,{\cal W})$ contains ${\cal O}(M)$ variables, whereas the corresponding subproblems in \cite{OURFDPAPER}  contain only ${\cal O}(K)$ and ${\cal O}(N)$ variables, respectively. Also, the iterative bandwidth fitting algorithm in \cite{OURFDPAPER} does not require the calculation of the dual variable $\alpha^k$ and the intermediate variables $X_m(\alpha^k,0)$, providing better computational efficiency. 
\end{remark}

\section{Multiple Non-Orthogonal Broadcast Channels}

\subsection{Problem Formulation}
We consider a system with multiple non-orthogonal broadcast channels, where each transmitter communicates with all its receivers on the same (assigned) frequency band at the same time. Denoting $X_{mki}$ as the symbol sent for receiver $m$ at instant $i$ in slot $k$, the signal received at receiver $m$ is ${Y}_{mki} = h_{mk}{X}_{mki}+\left(h_{mk}\sum_{m_0\neq m} {X}_{m_0ki} + {Z}_{mki}\right)$, where $h_{mk}$ represents the complex channel gain for receiver $m$ in slot $k$ and $Z_{mki}\sim {\sf CN}(0,1)$ is the i.i.d. complex Gaussian noise. We note that, $\sum_{m_0\neq m} {X}_{m_0ki}$ represents the interference and is treated as noise by receiver $m$. Moreover, we denote the channel gain and the energy consumption in each slot $k$ as $H_m^k\triangleq |h_{mk}|^2$ and $p_m^k \triangleq \frac{1}{T_c}\sum_i|X_{mki}|^2$, respectively.

We denote ${\tilde a}_{n}^k$ as the amount of bandwidth used by transmitter $n$. Then, we use the upper bound of the achievable rate over a weighted sum of the $M$ receivers and $K$ slots as the performance metric, given by \cite{IT}
\begin{equation}
 {\tilde C}_{\cal W}({\cal P},\tilde{\cal A}) \triangleq \sum_{n\in{\cal N}} \sum_{k\in{\cal K}} {\tilde a}_{n}^k \sum_{m\in{\cal M}_n} W_m  \log\left(1+\frac{p_m^kH_m^k/{\tilde a}_{n}^k}{\sum_{m_0\;|\; H_{m}^k< H_{m_0}^k}p_{m_0}^kH_m^k/{\tilde a}_{n}^k+1}\right)\ , 
\end{equation}
where $\tilde{\cal A}\triangleq \{{\tilde a}_n^k,\forall n\in{\cal N},k\in{\cal K}\}$. Note that, the rate in each slot is achieved by decoding the messages in the order of the channel quality \cite{FWC}, i.e., we decode the message from a weaker channel prior to that from a stronger channel. Moreover, we assume no two channels have the same gain in the same slot.

We define the energy-bandwidth allocation problem in multiple non-orthogonal broadcast channels as follows:
\begin{equation}\label{eq:weighedproblem}
{\tilde {\sf P}}_{\cal W}(\epsilon):\quad \max_{{\cal P},\tilde{\cal A}} {\tilde C}_{\cal W}({\cal P},\tilde{\cal A}) 
\end{equation}
subject to \eqref{eq:cst}, where $\sum_{m\in{\cal M}}a_m^k=1$ and ${a}_{m}^k \geq \epsilon$ is replaced by $\sum_{n}{\tilde a}_n^k = 1$ and ${\tilde a}_{n}^k \geq \epsilon$, respectively.

We note that, the above problem is non-convex due to the non-convexity of the objective function. To obtain the energy-bandwidth allocation, we first define ${\tilde p}_n^k \triangleq \sum_{m\in{\cal M}_n} p_m^k$ for all $n\in{\cal N}$ and rewrite \eqref{eq:weighedproblem} as
\begin{equation}\label{eq:noobj}
\max_{{\tilde p}_n^k, {\tilde a}_{m}^k} \left\{  \sum_{n} \sum_{k} \max_{\sum_{m\in{\cal M}_n} p_m^k={\tilde p}_m^k} \left\{{\tilde a}_{n}^k \sum_{m\in{\cal M}_n} W_m  \log\left(1+\frac{p_m^kH_m^k/{\tilde a}_{n}^k}{\sum_{m_0\;|\; H_{m}^k< H_{m_0}^k}p_{m_0}^kH_m^k/{\tilde a}_{n}^k+1}\right)\right\} \right\}\ .
\end{equation}

Denoting 
\begin{equation}\label{eq:esp}
 F_{n}^k(p) \triangleq \max_{\pi_m\;:\;\sum_{m\in{\cal M}_n} \pi_{m}=1,\pi_{m}\geq 0}\sum_{m\in{\cal M}_n}W_m  \log\left(1+\frac{\pi_m pH_m^k}{ \left(\sum_{m_0\;|\; H_{m}< H_{m_0}}\pi_m\right) pH_m^k+1}\right)\ ,
\end{equation}
we further write \eqref{eq:noobj} as
\begin{equation}\label{eq:nocvxobj}
\max_{{\cal P},\tilde{\cal A}}  {\tilde C}_{\cal W}({\cal P},\tilde{\cal A})  = \max_{{\tilde p}_n^k, {\tilde a}_{m}^k} \sum_{n} \sum_{k}{\tilde a}_{n}^k F_n^k({\tilde p}_n^k/{\tilde a}_{n}^k)\ ,
\end{equation}
where ${\tilde{\cal P}}\triangleq\{{\tilde p}_n^k,\forall n\in{\cal N},k\in{\cal K}\}$ is the total energy allocation.

To solve ${\tilde {\sf P}}_{\cal W}(\epsilon)$, we first solve \eqref{eq:nocvxobj} to obtain the optimal bandwidth allocation $\tilde{\cal A}$ and the optimal total energy allocation ${\tilde{\cal P}}$. Then, given the total energy allocation ${\tilde{\cal P}}$, we further optimally split the total energy for each receiver by solving \eqref{eq:esp}. 

The optimal solution to \eqref{eq:esp} is given in \cite{BEHRT}, which is summarized in the following Lemma:
\begin{lemma}\label{lm:splitting}
For any $(n,k)$, we have a set of energy cut-off lines $\{L_m^k,\forall m\in{\cal M}_n\}$ sorting in ascending order such that $L_a^k\leq L_b^k$ if $H_{a}^k > H_{b}^k$ for all $a,b\in{\cal M}_n$. For any $a\in{\cal M}_n$, the optimal energy splitting is
\begin{equation}\label{eq:splitting}
{p}_{a}^k=\left\{
\begin{array}{ll}
L_b^k-L_a^k,&\textrm{ if } L_b^k <  {\tilde p}_n^k \\
{\tilde p}_n^k -  L_b^k,&\textrm{ if } L_a^k \leq {\tilde p}_n^k \leq L_b^k  \\
0&\textrm{ if }  {\tilde p}_n^k < L_a^k  \\
\end{array}\right.\ ,
\end{equation}
where $L_a^k \leq L_b^k$ are two adjacent cut-off lines.
\end{lemma}

The procedure for computing  $\{L_m^k,\forall m\in{\cal M}_n\}$ is also given in \cite{BEHRT}.
 
\subsection{Solving the Problem in \eqref{eq:nocvxobj}}
The convexity of  $F_{n}^k(p)$ has been shown in \cite{BEHRT}, given by the following lemma:
\begin{lemma}\label{lm:concave}
$F_{n}^k(p)$ is strictly concave with respect to $p$, whose first-order derivative is continuous. 
\end{lemma}

Then, the problem in \eqref{eq:nocvxobj} is still an energy-bandwidth allocation problem with the rate function defined in \eqref{eq:esp}, which is increasing and jointly concave with respect to the total energy and bandwidth allocations. Note that the problem in \eqref{eq:nocvxobj} and the problem in \cite[Eqn. (9)-(10)]{OURFDPAPER} have the same feasible domain and the corresponding optimal energy allocations both follow the water-filling formula (will be shown later in this section). Then, it is easy to verify that the optimal energy discharge given by \cite[Eqn. (11)]{OURFDPAPER} and the iterative algorithm in \cite[Algorithm 1]{OURFDPAPER} can also give the optimal solution to the problem in \eqref{eq:nocvxobj}.

Hence we focus on the energy and bandwidth allocation subproblems as follows:
\begin{itemize}\item Energy allocation subproblem: Denote $\tilde{\cal A}_n\triangleq \{\tilde{a}_n^k,k\in{\cal K}\}$,
\begin{equation}
\tilde{\sf EP}_{n}(\tilde{\cal A}_n,{\cal W}):\quad\max_{{\tilde p}_n^k}   \sum_{n} \sum_{k}{\tilde a}_{n}^k F_n^k({\tilde p}_n^k/{\tilde a}_{n}^k)\ ,
\end{equation}
\begin{equation}
\textrm{subject to }\left\{
\begin{array}{l}
\tilde{E}_n^k - B_n^{\max} \leq \sum_{\kappa=1}^{k}{\tilde p}_{n}^{\kappa} \leq \tilde{E}_n^k \\
0\leq {\tilde p}_{n}^k\leq P_n\\
\end{array}\right. ,\ k\in{\cal K}.
\end{equation}
\item Bandwidth allocation subproblem: Denote  $\tilde{\cal P}_k\triangleq \{\tilde{p}_n^k,n\in{\cal N}\}$, 
\begin{equation}
\tilde{\sf BP}_k(\tilde{\cal P}_k,\epsilon,{\cal W}):\quad\max_{{\tilde a}_{n}^k}   \sum_{n} \sum_{k}{\tilde a}_n^k F_n^k({\tilde p}_n^k/{\tilde a}_n^k)\ ,
\end{equation}
subject to
\begin{equation}\label{eq:nobpcst}
\textrm{subject to }\left\{
\begin{array}{l}
\sum_{n=1}^N {\tilde a}_{n}^k \leq 1\\
{\tilde a}_n^k \geq \epsilon
\end{array}\right.,\ n\in{\cal N}.
\end{equation}
\end{itemize}

Using the Lagrangian multiplier defined in \eqref{eq:multiplier}, we first write the Lagrangian function for the problem in \eqref{eq:nocvxobj} as
\begin{equation}\label{L:I}
{\cal L}_N\triangleq \sum_{n} \sum_{k}{\tilde a}_{n}^k F_n^k({\tilde p}_n^k/{\tilde a}_{n}^k)+ {\cal M}(\tilde{\cal P},\tilde{\cal A})\ .
\end{equation}

\subsubsection{Solving the Energy Allocation Subproblem}
Since $\tilde{\sf EP}_{n}(\tilde{\cal A}_n,{\cal W})$ is a convex optimization problem with linear constraints, its K.K.T. conditions are sufficient and necessary for optimality when $\epsilon >0$ \cite{CO}. With ${\cal L}_{N}$ defined in \eqref{L:I}, we can write the first-order condition for the non-orthogonal broadcast channel as
\begin{equation}\label{eq:nob1o}
\partial \left({\tilde a}_{n}^kF_n^k({\tilde p}_{n}^k/{\tilde a}_{n}^k)\right)/\partial {\tilde p}_{n}^k\triangleq {(F_n^k)}'({\tilde p}_{n}^k/{\tilde a}_{n}^k) =  v_n^k - u_n^k
\end{equation}
where $v_n^k$ and $u_n^k$ are defined in \eqref{eq:level}, and $(F_n^k)'(p)$ denotes the first-order derivative of $F_n^k(p)$. For all $p \geq 0$, we further derive the derivative of $F_n^k(p)$ in closed-form:
\begin{proposition}\label{pp:derivative}
For any $p\geq 0$, the derivative of $F_n^k(p)$ is
\begin{equation}\label{eq:derivative}
{(F_n^k)}'(p)  = \max_{m\in{\cal M}_n}\left\{\frac{W_m}{p + 1/H_m^k}\right\}\ .
\end{equation}
\end{proposition}

The proof of Proposition \ref{pp:derivative} is provided in Appendix A.

Moreover, we note that ${(F_n^k)}'({\tilde p}_{n}^k/{\tilde a}_{n}^k) $ is strictly decreasing with respect to ${\tilde p}_{n}^k$ due to the strict concavity of $F_n^k(p)$. Then using \eqref{eq:nob1o} and Proposition \ref{pp:derivative}, ${\tilde p}_{n}^k$ can be uniquely determined as follows
\begin{align}
{\tilde p}_{n}^k &= {\tilde a}_n^k {\left({(F_n^k)}'\right)}^{-1}(1/w_n^k)\\
&=\min\left\{P_n,{\tilde a}_n^k\max_{m\in{\cal M}_n} \left\{\left[ {W_m}{w_n^k} - \frac{1}{H_m^k}\right]^+ \right\}\right\}\label{eq:nowf}
\end{align}
where $w_n^k = 1/(v_n^k - u_n^k)$ and $(\cdot)^{-1}$ denotes the inverse function.

We note that, since ${{\sf P}}_{\cal W}(\epsilon)$ and ${\tilde {\sf P}}_{\cal W}(\epsilon)$ have the same Lagrangian multipliers, by analyzing the K.K.T. conditions and using Proposition \ref{pp:derivative}, it is easy to verify that the changes of $w_n^k$ still follows Proposition \ref{pp:wf}, i.e., it may only increase/decrease at the BDP/BFP. Then, we treat \eqref{eq:nowf} as a water-filling formula and the water level is determined by
\begin{equation}\label{eq:nowft}
\sum_{k=a+1}^b {\tilde p}_n^k(w^{ab})= E^b - E^a + \left(\mathbb{I}(a \textrm{ is BFP}) - \mathbb{I}(a \textrm{ is BDP}) \right) B_n^{\max}
\end{equation}
where ${\tilde p}_n^k(w^{ab})$ is calculated by \eqref{eq:nowf} with $w_n^k=w^{ab}$ for $k\in[a+1,b]$.

As for the energy allocation problem in multiple orthogonal broadcast channels, since here the water level change also occurs at BDP/BFPs, we can use the water-filling in \eqref{eq:nowf}-\eqref{eq:nowft} to replace the conventional water-filling operation in \cite[Algorithm 2]{2014arXiv1401.2376W}, and then the BDP/BFP set can be obtained. After obtaining the BDP/BFP set, using \eqref{eq:nowf}-\eqref{eq:nowft}, we obtain the optimal total energy allocation.

\subsubsection{Solving the Bandwidth Allocation Subproblem} When $\sum_{n\in{\cal N}}\tilde{p}_n^k=0$, the sum-rate in slot $k$ is zero. Thus, in this subsection we focus on the case $\sum_{n\in{\cal N}}\tilde{p}_n^k>0$.  

Since $\tilde{\sf BP}_k(\tilde{\cal P}_k,\epsilon,{\cal W})$ is a convex optimization problem with linear constraints, its K.K.T. conditions are sufficient and necessary for optimality when $\epsilon >0$ \cite{CO}. The first-order condition can be written as 
\begin{equation}\label{kkt:noal}
\frac{\partial \left({\tilde a}_n^kF_n^k({\tilde p}_n^k/{\tilde a}_n^k)\right)}{\partial {\tilde a}_n^k} = F_n^k({{\tilde p}_n^k}/{{\tilde a}_n^k})  -  {(F_n^k)}'({{\tilde p}_n^k}/{{\tilde a}_n^k})  {{\tilde p}_n^k}/{{\tilde a}_n^k} = \alpha^k\ ,\ n\in{\cal N},k\in{\cal K}\ ,
\end{equation}
where the value of $F_n^k({{\tilde p}_n^k}/{{\tilde a}_n^k}) $ can be calculated using the algorithm in \cite{BEHRT}. Taking the constraints in \eqref{eq:nobpcst} into account, ${\tilde a}_{n}^k$ must satisfy
\begin{equation}\label{eq:cstbd}
\sum_{n=1}^N \max\{{\tilde a}_n^k,\epsilon\} = 1,\ k\in{\cal K}\ .
\end{equation}
We note that, for each $k\in{\cal K}$, we have $N+1$ equations [\eqref{kkt:noal} for all $n\in{\cal N}$ and \eqref{eq:cstbd}] and $N+1$ variables [${\tilde a}_{n}^k$ for all $ n\in{\cal N}$ and $\alpha^k$]. Therefore, all the variables ${\tilde a}_{n}^k$ can be uniquely determined by solving the equation group given $k\in{\cal K}$.

Since $F_n^k(p)$ is concave by Lemma \ref{lm:concave}, $aF_n^k(p/a)$ is jointly concave with respect to $p$ and $a$. Then, ${\partial \left(aF_n^k(p/a)\right)}/{\partial a} $ is non-increasing with respect to $a$ given $p$. Also, the left-hand-side of \eqref{eq:cstbd} is non-decreasing with respect to $a$. Therefore, given $\alpha^k$, we can use the bisection method to find the corresponding ${\tilde a}_{n}^k(\alpha^k)$ in \eqref{kkt:noal}. Finally we can use the bisection method again to determine the proper $\alpha^k$ such that \eqref{eq:cstbd} is satisfied. The procedure for computing the bandwidth allocation is summarized as follows.\\
\begin{minipage}[h]{6.5 in}
\rule{\linewidth}{0.3mm}\vspace{-.05in}
{\bf {\footnotesize Algorithm 2 - Solving bandwidth allocation subproblem $\tilde{\sf BP}_k({\cal P}_k,\epsilon,{\cal W})$}}\vspace{-.1in}\\
\rule{\linewidth}{0.2mm}
{ {\small
\begin{tabular}{ll}
	\;1:&  Initialization\\
	\;& Specify initial $\alpha_u^k>\alpha_l^k > 0$ such that $\sum_{n=1}^N \max\{{\tilde a}_n^k(\alpha_u^k),\epsilon\}<1<\sum_{n=1}^N \max\{{\tilde a}_n^k(\alpha_l^k),\epsilon\}$ \\
	\;& Specify error tolerance $\delta>0$\\
    \;2:& {\bf REPEAT}\\
    \;&\quad $\alpha \leftarrow (\alpha_u^k+\alpha_l^k)/2$\\
    \;& \quad {\bf FOR} all $n\in{\cal N}$\\
      \;(*)& \quad\quad Solve \eqref{kkt:noal} to obtain ${\tilde a}_n^k(\alpha)$ using the bisection method\\
       \;& \quad {\bf ENDFOR}\\
      	\;&\quad {\bf IF} $|\sum_{n=1}^N \max\{{\tilde a}_n^k(\alpha^k),\epsilon\}-1| < \delta$ {\bf THEN} Goto step 4 {\bf ENDIF}\\
      	\;& \quad {\bf IF} $\sum_{n=1}^N \max\{{\tilde a}_n^k(\alpha^k),\epsilon\} > 1$ {\bf THEN} $\alpha_l^k\leftarrow \alpha$ {\bf ELSE}  $\alpha_h^k\leftarrow \alpha$ {\bf ENDIF}\\
      	      	\;3:& {\bf FOR} all $n\in{\cal N}$\\
      	    \;&  \quad Calculate ${\tilde a}_{n}^k$ by \eqref{eq:oba}  \\
      	\;& {\bf ENDFOR}\\
\end{tabular}}}\\
\rule{\linewidth}{0.3mm}
\end{minipage}\vspace{.2 in}

The complexity of Algorithm 2 is ${\cal O}(N)$.

\begin{remark}
Comparing Algorithm 2 with Algorithm 1, the main difference lies in the step marked by ``*'', where the corresponding bandwidth allocations $a_m^k$ and ${\tilde a}_n^k$ are calculated by solving the same equation [i.e., \eqref{eq:equation}] in Algorithm 1 and multiple different equations [i.e., \eqref{kkt:noal} with different $F_n^k({\tilde p}_n^k/{\tilde a}_n^k)$ for all $n\in{\cal N}$] in Algorithm 2.
\end{remark}

\subsection{Special Case: Equal Weights}

When $W_m=1$ for all $m\in{\cal M}$, by Proposition \ref{pp:derivative}, we have
\begin{equation}
{(F_n^k)}'(p)  = \max_{m\in{\cal M}_n}\left\{\frac{1}{p + 1/H_m^k}\right\}\ ,
\end{equation}
for all $p\geq 0$.
Since, given any $a,b\in{\cal M}_n$ such that $H_a > H_b > 0$, we have $1/(p + 1/H_a) > 1/(p + 1/H_b)$ for all $p \geq 0$, then we have
\begin{equation}
{(F_n^k)}'(p) = \max_{m\in{\cal M}_n}\left\{\frac{1}{p + 1/H_m^k}\right\} = \frac{1}{p + 1/\max_{m\in{\cal M}_n}\{H_m^k\}}\ .
\end{equation}
Therefore, by \eqref{eq:esp}, we must have 
\begin{equation}
F_n^k(p) = \log(1 + pH_{m_n^k}^k)
\end{equation}
where $m_n^k\triangleq \arg\max_{m\in{\cal M}_n}\{H_m^k\}$, i.e.,  each transmitter  uses only the strongest channel to transmit in each slot. Then, we have the following corollary.

\begin{corollary}\label{cl:ew}
Theorem \ref{thm:ew} also holds for the network with multiple non-orthogonal broadcast channels. Moreover, with equal weights,  networks with multiple orthogonal and non-orthogonal broadcast channels achieve the same maximum throughput.
\end{corollary}

\begin{remark}
When the weights are equal, by Corollary \ref{cl:ew}, the energy-bandwidth allocation for multiple  orthogonal broadcast channels is equivalent to that for multiple point-to-point channels treated in \cite{OURFDPAPER}. Comparing to the algorithms in \cite{OURFDPAPER}, the general algorithms in this section involve solving subproblems with more variables and constraints and the additional  calculations of $F_n^k(p)$ and $\alpha$. Thus we should use Corollary \ref{cl:ew} along with the algorithms in \cite{OURFDPAPER} to solve the energy allocation problem when the weights are equal.
\end{remark}

\subsection{Achievable Rate Regions}
Denoting $C_{O,m}({\cal P},{\cal A})$ and $C_{N,m}({\cal P},{\cal A})$ as the sum-rate of receiver $m$ achieved by the energy-bandwidth allocation $({\cal P,A})$ in $K$ slots for multiple orthogonal and non-orthogonal broadcast channels, respectively. Then, the rate region can be defined as ${\cal R}_{(\cdot)}\triangleq \{(r_{1},r_{2},\ldots,r_{M})\ |\  0\leq r_m \leq C_{(\cdot),m}({\cal P,A}),\ {\cal P,A} \textrm{ are feasible}\}$, where $(r_{1},r_{2},\ldots,r_{M})$ is the sum-rate vector for all receivers.

\begin{lemma}\label{eq:cvxregion}
The rate region ${\cal R}_O$ is strictly convex for the network with multiple orthogonal broadcast channels.
\end{lemma}
\begin{IEEEproof}
Consider two sum-rate vectors $R^1,R^2\in{\cal R}_O$ and the corresponding energy-bandwidth allocation as $({\cal P}^1,{\cal A}^1)$ and $({\cal P}^2,{\cal A}^2)$. Then, given any $\theta\in(0,1)$ and $\bar{\theta} = 1 - \theta$, consider $R^3 = \theta  R^1+\bar{\theta}  R^2$, where $R^i\triangleq (r_{1}^i,r_{2}^i,\ldots,r_{M}^i)$. We note that, $C_{O,m}({\cal P},{\cal A})$ is sum of a series of log functions which are strictly concave with respect to $p_m^k$ and $a_m^k$. Then, for $m\in{\cal M}$, we have
\begin{align}
r^3_{m} &= \theta  r_{m}^1+ \bar{\theta} r_{m}^2\\
&\leq \theta  C_{O,m}({\cal P}^1,{\cal A}^1)+ \bar{\theta} C_{O,m}({\cal P}^2,{\cal A}^2)\\
&<  C_{O,m}(\theta {\cal P}^1+ \bar{\theta} {\cal P}^2,\theta {\cal A}^1+ \bar{\theta} {\cal A}^2)
\end{align}
where ${\cal P}^3\triangleq \theta {\cal P}^1+ \bar{\theta} {\cal P}^2$ and ${\cal A}^3\triangleq \theta {\cal A}^1+ \bar{\theta} {\cal A}^2$. Note that, since ${\sf P}_{\cal W}(\epsilon)$ is a convex optimization problem and its feasible domain is also convex, $({\cal P}^3,{\cal A}^3)$ is a feasible energy-bandwidth allocation. Then, by definition we have $R^3\in{\cal R}_{O}$ and thus ${\cal R}_O$ is a strictly convex set.
\end{IEEEproof}

Moreover, for the network with multiple non-orthogonal broadcast channels, we define a convex region
\begin{equation}
\bar{\cal R}_N \triangleq \left\{(r_{1},r_{2},\ldots,r_{M}):r_m \leq{\tilde {\sf P}}_{\{W_m=1,W_i=0,\forall i\neq m\}}(0), \sum_{m}r_m \leq {\tilde {\sf P}}_{\{W_m=1,\forall m\}}(0) \right\}\ .
\end{equation}
Note that for ${\cal W}=\{W_m=1,W_i=0,\forall i\neq m\}$, $\tilde{\sf P}_{\cal W}(0)$ and ${\sf P}_{\cal W}(0) $ maximize the sum-rate for the single receiver $m$ and the two problems are the same. Then we have
\begin{equation}
\tilde{\sf P}_{\{W_m=1,W_i=0,\forall i\neq m\}}(0) = {\sf P}_{\{W_m=1,W_i=0,\forall i\neq m\}}(0)  = \max_{{\cal P,A}\textrm{ are feasible }} C_{(\cdot),m}({\cal P,A}),\ m\in{\cal M} \ .
\end{equation}
For $W_m=1, m\in{\cal M}$, by Theorem \ref{thm:ew} and Corollary \ref{cl:ew}, $\tilde{\sf P}_{\{W_m=1,\forall m\}}(0)$ and ${\sf P}_{\{W_m=1,\forall m\}}(0)$ have the same solution, which can be denoted as $({\cal P}^*,{\cal A}^*)$. For any $(r_1,r_2,\ldots,r_M)\in{\cal R}_{N}$, by definition, we have $r_m \leq \max_{{\cal P,A}\textrm{ are feasible }} C_{N,m}({\cal P,A})$ and  $ \sum_{m}r_m \leq \sum_{m}C_{N,m}({\cal P}^*,{\cal A}^*)$. Then, we have ${\cal R}_N\subseteq \bar{\cal R}_N$ and the sum-rate vectors $(C_{(\cdot),1}({\cal P}^*,{\cal A}^*),C_{(\cdot),2}({\cal P}^*,{\cal A}^*), \ldots, C_{(\cdot),M}({\cal P}^*,{\cal A}^*))$ and $(\ldots,0,{\sf P}_{\{W_m=1,W_i=0,\forall i\neq m\}}(0),0,\ldots)$ for all $m\in{\cal M}$ can be achieved with both orthogonal and non-orthogonal broadcast.

\begin{figure}
\centering
\includegraphics[width=0.5\textwidth]{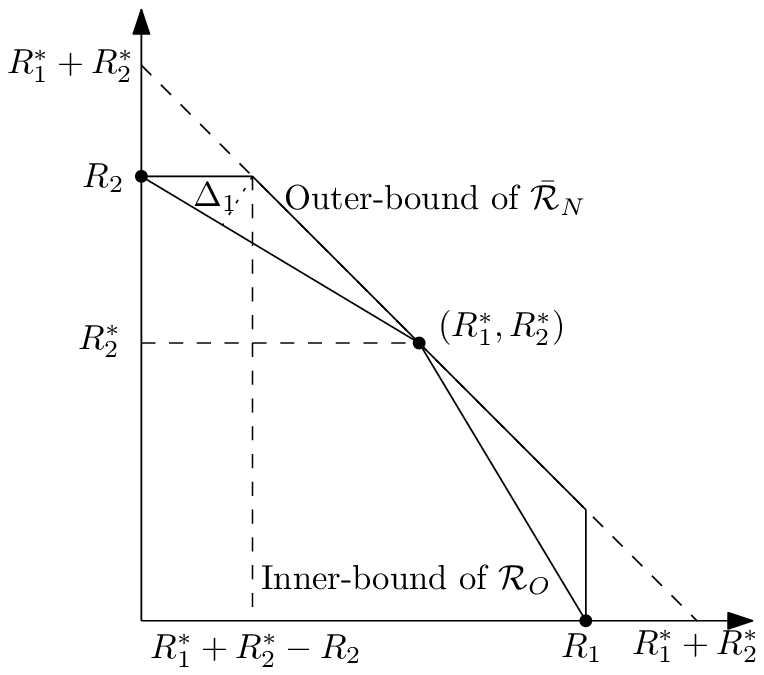}
\caption{Rate regions of orthogonal and non-orthogonal broadcast channels.}
\label{fg:GF}
\end{figure}

We give an example for the network with one transmitter and two receivers. According to the above analysis, ${\cal R}_O$ and $\bar{\cal R}_N$ have three common points on the boundary as shown in Fig. \ref{fg:GF}: $(R_1,0)$ for $\{W_{11}=1,W_{12}=0\}$, $(0,R_2)$ for $\{W_{11}=0,W_{12}=1\}$, and $(R^*_1,R^*_2)$ for $\{W_{11}=W_{12}=0.5\}$. Due to the concavity of  ${\cal R}_O$ and $\bar{\cal R}_N$, the maximum improvement (Euclidean distance between boundary of ${\cal R}_O$ and $\bar{\cal R}_N$) of using the non-orthogonal broadcast channel is bounded by
\begin{equation}
\Delta= \max\left\{\frac{(R_2-R^*_2)(R^*_1+R^*_2-R_2)}{\sqrt{(R_2 - R^*_2)^2+{R^*_1}^2}},\frac{(R_1-R^*_1)(R^*_2+R^*_1-R_1)}{\sqrt{(R_1 - R^*_1)^2+{R^*_2}^2}}\right\}\ .
\end{equation}

\section{Achieving Proportional Fairness in Orthogonal Broadcast Channels}
In this section, we formulate a proportionally-fair (PF) throughput maximization problem for the network with multiple orthogonal broadcast channels, and show that it can be converted to a weighted throughput  maximization problem with some proper weights.

\subsection{PF Throughput Maximization}
We consider the following utility function
\begin{equation}
U({\cal P},{\cal A}) \triangleq \sum_{m\in{\cal M}}\log\left(\sum_{k\in{\cal K}} a_m^k \log(1+\frac{p_m^kH_m^k}{a_m^k})\right)
\end{equation}
Then, the PF throughput maximization problem is formulated as
\begin{equation}
{\sf F}_{\epsilon}:\quad \max_{{\cal P},{\cal A}} U({\cal P},{\cal A})
\end{equation}
subject to the constraints in \eqref{eq:cst}, whose solution is known to result in proportional fairness \cite{PFRAEH}\cite{CPFSAG}. Without loss of generality, we assume $\tilde{E}_n^{K} > 0$ for all $n\in{\cal N}$ and thus each transmitter achieves a non-zero sum-rate to make the PF throughput lower bounded.

We next convert ${\sf F}_{\epsilon}$ into a weighted throughput problem ${\sf P}_{\cal W}(\epsilon)$. Specifically, given $\cal W$, we denote $R_{m}({\cal W})$ as the sum-rate achieved for receiver $m$ by the optimal solution  to ${\sf P}_{\cal W}(\epsilon)$; we also denote $\bar{R}_{m}$ as the sum-rate achieved for receiver $m$ by the optimal solution to  ${\sf F}_{\epsilon}$.
We note that, since the rate region ${\cal R}_O$ is strictly convex, $R_{m}({\cal W})$, which is the tangent point of a hyperplane (defined by $\cal W$) to ${\cal R}_O$, is continuous in ${\cal W}$.  
\begin{theorem}\label{thm:pfeq}
Given $\cal W$, the optimal solution to ${\sf P}_{\cal W}(\epsilon)$ is also optimal to ${\sf F}_\epsilon$, if and only if, there exists $\theta > 0$ such that $W_m R_{m}({\cal W}) = \theta$ for all $m\in{\cal M}$, where  $R_{m}({\cal W})$ is the sum-rate achieved for receiver $m$ by the optimal solution  to ${\sf P}_{\cal W}(\epsilon)$.
\end{theorem}
\begin{IEEEproof}
We note that ${\sf P}_{\cal W}(\epsilon)$ and ${\sf F}_\epsilon$ have the same decision variables and the same constraints and they can use the same Lagrangian multiplier as defined in \eqref{eq:multiplier}. Then, the Lagrangian functions for  ${\sf P}_{\cal W}(\epsilon)$ and ${\sf F}_\epsilon$ can be defined as \eqref{L:O} and 
\begin{align}
{\cal L}_F\triangleq &\sum_{m\in{\cal M}}\log\left(\sum_{k\in{\cal K}} a_m^k \log(1+\frac{p_m^kH_m^k}{a_m^k})\right)+{\cal M}({\cal P,A}),
\end{align}
respectively. Taking the first-order derivatives with respect to $p_m^k$, we have
\begin{align}
&\frac{\partial {\cal L}_P}{\partial p_m^k} = W_m \frac{\partial\left(a_m^k \log(1+\frac{p_m^kH_m^k}{a_m^k})\right)}{\partial p_m^k} + \frac{\partial {\cal M}}{\partial p_m^k}\ , \\
&\frac{\partial {\cal L}_F}{\partial p_m^k} =\frac{1}{\bar{R}_{m}}  \frac{\partial \left(a_m^k \log(1+\frac{p_m^kH_m^k}{a_m^k})\right)}{\partial p_m^k} + \frac{\partial {\cal M}}{\partial p_m^k}\ ;
\end{align}
also, we can obtain the derivative with respect to $a_{m}^k$ in the same form as above. Note that, for ${\sf P}_{\cal W}(\epsilon)$ and ${\sf F}_\epsilon$, their K.K.T. conditions are sufficient and necessary for optimality when $\epsilon > 0$. Also, since $\bar{R}_m$ is the sum-rate achieved for receiver $m$ by the optimal solution  to ${\sf F}_\epsilon$ and $R_m({\cal W})$ is the sum-rate achieved by the optimal solution to ${\sf P}_{\cal W}(\epsilon)$, when $W_m = 1/ R_{m}({\cal W})$ for all $m\in{\cal M}$, the solution satisfies the K.K.T. conditions of ${\sf F}_\epsilon$  also satisfies those of ${\sf P}_{\cal W}(\epsilon)$, and vice versa. Therefore, ${\sf P}_{\cal W}(\epsilon)$ and ${\sf F}_\epsilon$ have the same optimal solution. Moreover, we note that scaling $W_m$ by a positive factor $\theta$ does not affect the optimality of ${\sf P}_{\cal W}(\epsilon)$ and thus the above equivalence condition can be further relaxed to $W_m = \theta/ R_{m}({\cal W})$ where $\theta > 0$. Furthermore, since the objective functions of the two problems are both continuous, we can further extend the result to the case of $\epsilon=0$.
\end{IEEEproof}



We call $\cal W$ the {\em PF weights} if ${\sf P}_{\cal W}(\epsilon)$ and ${\sf F}_\epsilon$ have the same optimal solution.

\subsection{Obtaining the PF Weights}

To obtain the PF weights, we first define an optimization problem:
\begin{equation}\label{eq:subgproblem}
\min_{(\frac{1}{W_1},\frac{1}{W_2},\ldots,\frac{1}{W_M})\in{\cal R}_O}\max_{{\cal P,A}}\left\{\sum_{m}W_m\left(\sum_k a_m^k \log(1+\frac{p_m^kH_m^k}{a_m^k}) - 1/W_m\right) \right\}
\end{equation}
subject to
\begin{equation}\label{eq:subgcst}
\left\{
\begin{array}{ll}
\sum_{k} a_m^k \log(1+\frac{p_m^kH_m^k}{a_m^k}) \geq 1/W_m,& n\in{\cal N},m\in{\cal M}_n\\
\textrm{Constraints in } \eqref{eq:cst}&\\
\end{array}\right.\ .
\end{equation}
We note that, since $1/W_m$ is drawn from the rate region ${\cal R}_O$, the optimal value of \eqref{eq:subgproblem} is zero, and $\sum_k a_m^k \log(1+\frac{p_m^kH_m^k}{a_m^k}) - 1/W_m=0$. By Theorem \ref{thm:pfeq}, the PF weights is also the optimal solution to \eqref{eq:subgproblem}. 
Then, denoting $\bar{W}_{m}\triangleq W_m+\lambda_m$ where $\lambda_m \geq 0$ is the dual variable, we convert the inner maximization problem in \eqref{eq:subgproblem} to its dual problem and \eqref{eq:subgproblem} can be further written as
\begin{equation}\label{eq:subdual}
\min_{(\frac{1}{W_1},\frac{1}{W_2},\ldots,\frac{1}{W_M})\in{\cal R}_O,\bar{W}_{m}\geq W_m}\max_{{\cal P,A}\textrm{ subject to \eqref{eq:cst}}}\left\{\sum_{m}\bar{W}_{m}\left(\sum_k a_m^k \log(1+\frac{p_m^kH_m^k}{a_m^k}) - 1/W_m\right) \right\}\ .
\end{equation}

Note that the inner problem of \eqref{eq:subdual} is equivalent to the weighted throughput optimization problem ${\sf P}_{\bar{\cal W}}(\epsilon)$ with an additional constant term $\sum_m \bar{W}_{m}/W_m$, where $\bar{\cal W}=\{\bar{W}_m,m\in{\cal M}\}$. Thus, when $W_m=\bar{W}_m = 1/R_m({\cal W})$, the problem in \eqref{eq:subdual} is optimally solved (the optimal value is zero, which is same as the problem in \eqref{eq:subgproblem}) and by Theorem \ref{thm:pfeq} the optimal PF weights are obtained. Then, we can write the subgradient for the outer minimization problem in \eqref{eq:subdual} as \cite{CO}
\begin{align}
g_{\bar{W}_{m}} &=  R_{m}(\bar{\cal W}) - 1/W_m \ ,\label{sb:1}\\
g_{W_m} &= {{\bar{W}_{m}}}/{W_m^2} > 0\label{sb:2}\ .
\end{align}

Since the subgradient of $W_m$ is positive, the optimal $1/W_m$ is on the positive boundary of ${\cal R}_O$. Note that $(R_1(\bar{\cal W}),R_2(\bar{\cal W}),\ldots,R_M(\bar{\cal W}))$ is on the positive boundary of ${\cal R}_O$ and changes continuously as $\bar{W}$ changes. Then, the following update rule
\begin{equation}\label{eq:updateo}
\left\{\begin{array}{l}
{W}_m  \leftarrow \min\left\{\bar{W}_m,\left[{W}_m - \delta\cdot (R_{m}(\bar{\cal W}) - 1/W_m)\right]^+\right\}\\
\bar{W}_m  \leftarrow \max\left\{{W}_m, \left[\bar{W}_m - \delta\cdot g_{\bar{W}_m}\right]^+\right\}
\end{array}\right.\ ,
\end{equation}
enforces that $W_m$ always moves closer to the point on the positive boundary of ${\cal R}_O$ and $\bar{W}_m$ is updated by the subgradient. Specifically, if we fix $W_m$ (or $\bar{W}_m$) and update $\bar{W}_m$ (or $W_m$) only using the second (first) term in \eqref{eq:updateo},  $W_m$ (or $\bar{W}_m$) can converge and the optimal $\bar{W}_m$ (or $W_m$) can be obtained for the fixed $W_m$ (or $\bar{W}_m$).

To find the PF weights, we need to obtain the optimal solution to \eqref{eq:subdual} such that $W_m=\bar{W}_m$. Specifically, we choose the same initial condition and step size for $W_m$ and $\bar{W}_m$, and simultaneously update $W_m$ and $\bar{W}_m$ in each iteration. Then, $W_m$ and $\bar{W}_m$ remain the same in each iteration and the update rule becomes
\begin{equation}\label{eq:update}
W_m^{(i+1)} = \bar{W}_m^{(i+1)} \leftarrow \left[\bar{W}_m^{(i)} - \delta(i)\cdot g_{\bar{W}_m}^{(i)}\right]^+\ ,
\end{equation}
where the step size $\delta(i)$ satisfies $\lim_{i\rightarrow \infty} \delta(i) = 0$ and $\sum_{i=1}^{+\infty} \delta(i)=+\infty$, e.g., $\delta(i) = 1/i$. In particular, if $W_m^{(i+1)}$ can converge, the problem in \eqref{eq:subdual} is optimally solved and finally we have $\bar{W}_m=W_m$ for all $m\in{\cal M}$, i.e., $ R_{m}({\cal W}) = 1/W_m$. By Theorem \ref{thm:pfeq}, $\cal W$ are the PF weights. 

The procedure for computing the PF energy-bandwidth allocation is summarized as follows.\\
\begin{minipage}[h]{6.5 in}
\rule{\linewidth}{0.3mm}\vspace{-.05in}
{\bf {\footnotesize Algorithm 3 -  PF energy-bandwidth allocating algorithm}}\vspace{-.1in}\\
\rule{\linewidth}{0.2mm}
{ {\small
\begin{tabular}{ll}
	\;1:&  Initialization\\
	\;& $i=0$\\
	\;& Specify the initial fairness weights ${\cal W}^{(0)}$, convergence threshold $\delta_0$, maximum iteration number $I$\\
    \;2:& Obtaining the PF weight\\
    \;& {\bf REPEAT}\\
      \;& \quad $i\leftarrow i+1$\\
      	\;& \quad Solve ${\sf P}_{{\cal W}^{(i-1)}}(\epsilon)$ to obtain $({\cal P}^{(i)},{\cal A}^{(i)})$\\
      	\;& \quad Update ${\cal W}^{(i)}$ by \eqref{eq:update}\\
	\;& {\bf UNTIL}  $\sum_{m}|R_{m}({\cal W}^{(i)})  - 1/W_m^{(i)}| \leq \delta_0$ {\bf OR} $i= I$\\
	\;3:& Choose the energy-bandwidth Allocation\\
	\;& $({\cal P}^{(i)},{\cal A}^{(i)})$ is the obtained energy-bandwidth allocation
\end{tabular}}}\\
\rule{\linewidth}{0.3mm}
\end{minipage}\vspace{.05 in}

Note that, the convergence of the proposed algorithm is highly dependent on the selection of the initial value, i.e., ${\cal W}^{(0)}$. Specifically, we can set 
\begin{equation}\label{eq:appPFweight}
\frac{1}{W_m^{(0)}} \approx \mathbb{E}_{\{\tilde{E}_n^k,H_n^k\}}\left[\bar{R}_{m}(K)\right]\ ,
\end{equation}
as the initial PF weights, where $\bar{R}_{m}(K)$ denotes the sum-rate achieved by the solution to ${\sf F}_\epsilon$ given the realizations $\{\tilde{E}_n^k,H_n^k, n\in{\cal N},k\in{\cal K}\}$ in the scheduling period $K$, and the simulation results in Section VI demonstrate that the optimal performance is approached closely in a few iterations.

\section{Simulation Results}
\begin{figure}
\centering
\begin{minipage}[t]{0.49\textwidth}
\includegraphics[width=1\textwidth]{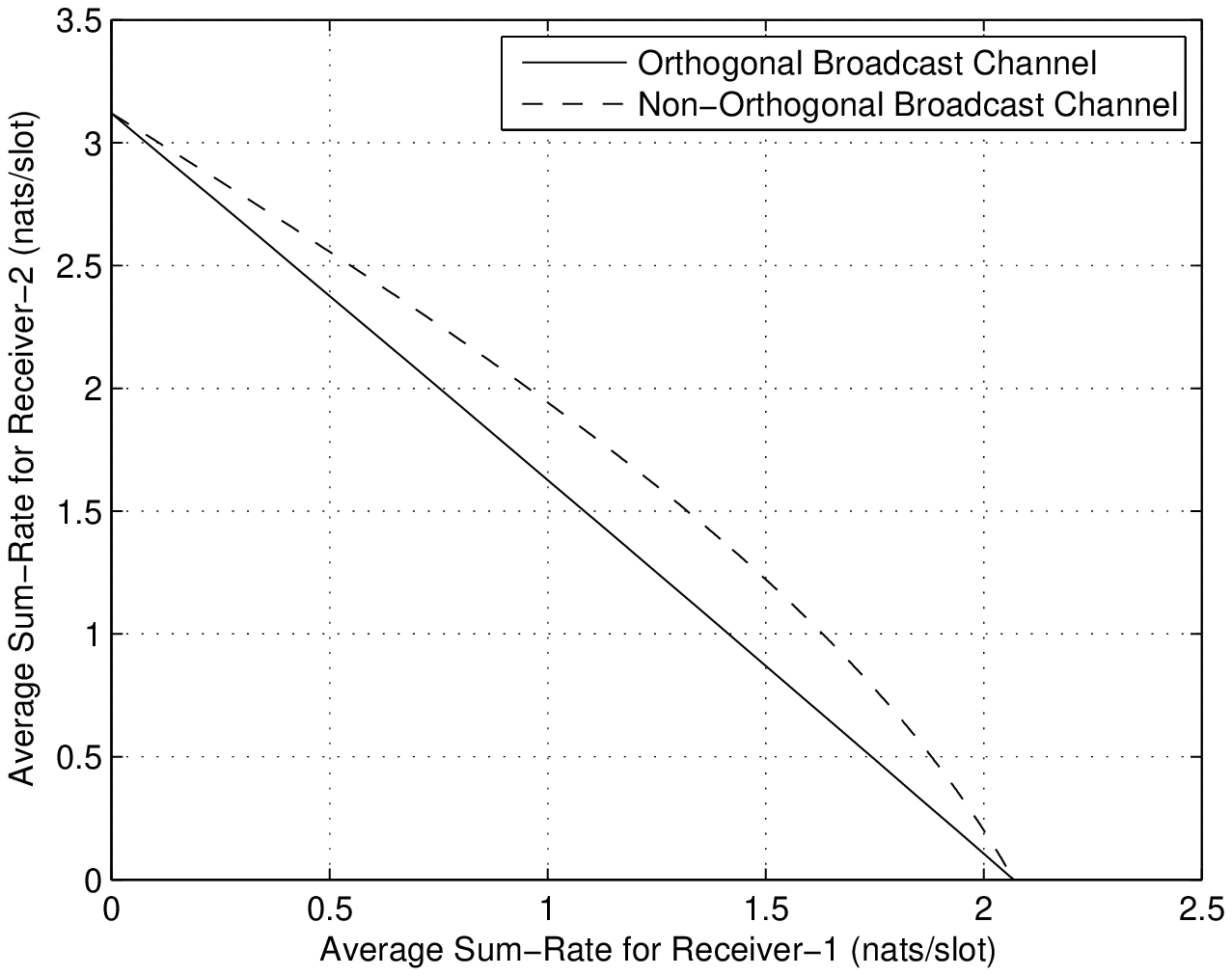}
\caption{Achievable sum-rate regions of two-user orthogonal/non-orthogonal broadcast channels ($K=1$).}
\label{fg:bc1}
\end{minipage}
\begin{minipage}[t]{0.49\textwidth}
\centering
\includegraphics[width=1\textwidth]{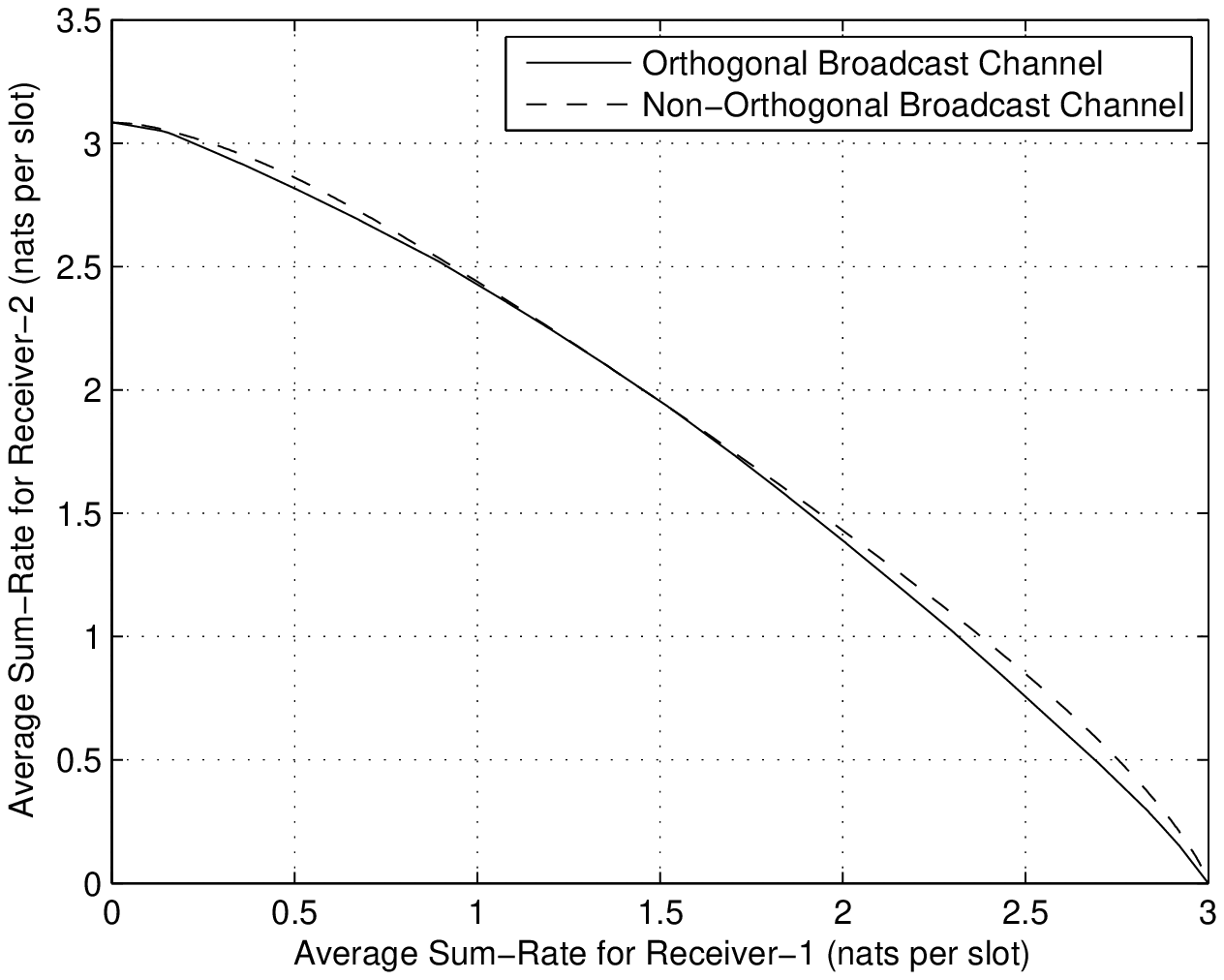}
\caption{Achievable sum-rate regions of two-user orthogonal/non-orthogonal broadcast channels ($K=10$).}
\label{fg:bc10}
\end{minipage}
\end{figure}

We first focus on a single transmitter and compare the achievable rate regions for orthogonal and non-orthogonal two-user broadcast channels, i.e, $N=1$  and $M=2$. For the transmitter, we set the  initial battery level $B_n^k=0$, the battery capacity $B_n^{\max}=20$ units, and we do not apply the maximum power constraint. We generate the realizations of the harvested energy $E_n^k$ and channel gains $H_m^k$ following the truncated Gaussian distribution ${\cal N}(10,2)$ and the Rayleigh distribution with the parameter $2$, respectively. Moreover, we consider two scheduling period, $K=1$ slot and $K=10$ slots, and show the sum-rate improvement by the non-orthogonal broadcast over the orthogonal broadcast in Fig. \ref{fg:bc1} and Fig.  \ref{fg:bc10}, respectively. Specifically, we note that when $K=10$ the improvement is quite marginal. Moreover, in Fig. \ref{fg:bc10}, two curves share  three common points corresponding to the sum-rate achieved by the solution to ${\cal P}_0(W_1,W_2)$ for $(W_1,W_2) = (1,0), (0.5,0.5)$ and $(0,1)$, respectively. Also, when $W_1=W_2=0.5$, the sum-rates are maximized for both the orthogonal and non-orthogonal broadcast, which are same.

\subsection{Weighted Sum-Rate Maximization}

We then consider a network with multiple broadcast channels where there are $N=3$ transmitters and each communicates with $2$ receivers, i.e., ${\cal M}_1=\{1,2\}, {\cal M}_2=\{3,4\}, {\cal M}_3=\{5,6\}$. We set the scheduling period as $K=20$ slots. For each transmitter $n$, we set the initial battery level $B_n^0=0$ and the battery capacity $B_n^{\max}=20$ units. We assume that the harvested energy $E_n^k$ follows a truncated Gaussian distribution with mean $\mu_{n}$ and variance of $2$. We also assume a Rayleigh fading  channel with the parameter $\sigma_{m}$. 

For comparison, we consider two simple scheduling strategies, namely, the {\em greedy energy policy} and the {\em equal bandwidth policy}. For the greedy energy policy, each transmitter first tries to use up the available energy in each slot. Then, given the available energy for each transmitter, we solve the energy-bandwidth allocation problem slot by slot, i.e., ${\sf P}_{\cal W}(0)$ for $K=1$, to calculate the energy and bandwidth allocated for each receiver. For the equal bandwidth policy, we first assign the bandwidth for each transmitter equally. Then, given the assigned bandwidth for each transmitter, we solve an energy-bandwidth allocation problem transmitter by transmitter, i.e., ${\sf P}_{\cal W}(0)$ for $N=1$,  to calculate the energy and bandwidth (for orthogonal broadcast channel only) allocated for each receiver.

\begin{figure}
\centering
\begin{minipage}[t]{0.49\textwidth}
\includegraphics[width=1\textwidth]{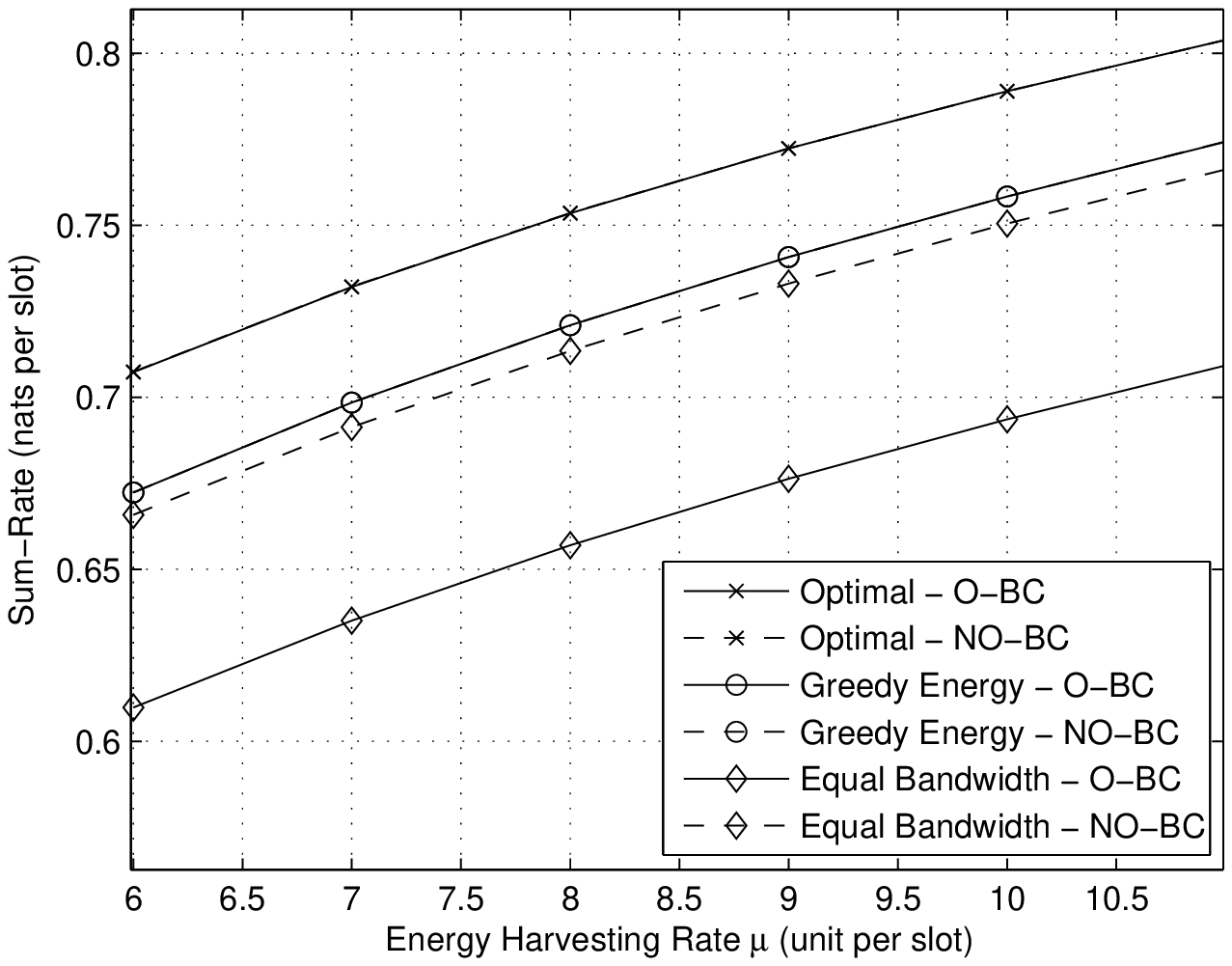}
\caption{ Sum-rate comparisons for different policies without the maximum power (${\cal W}_1$).}
\label{fg:ew}
\end{minipage}
\begin{minipage}[t]{0.49\textwidth}
\includegraphics[width=1\textwidth]{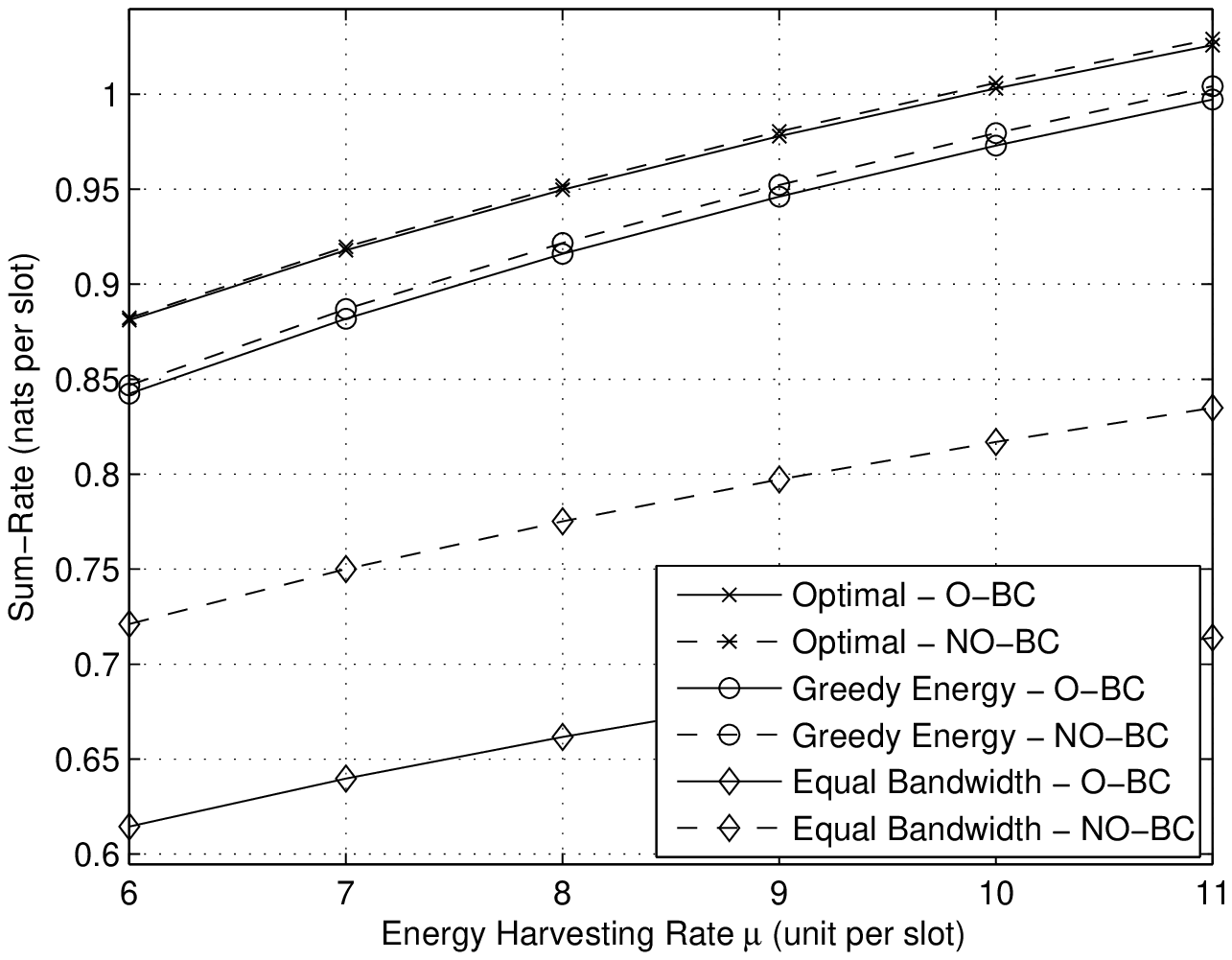}
\caption{Weighted sum-rate comparisons for different policies without the maximum power (${\cal W}_2$).}
\label{fg:new}
\end{minipage}
\end{figure}

To compare the performance of the different algorithms and policies, we evaluate the (weighted) sum-rate for the multiple orthogonal  broadcast channels (O-BCs) and non-orthogonal broadcast channels (NO-BCs), respectively.  We use  ${\cal W}_1 = \{W_m=1/6\}$ and ${\cal W}_2 = \{W_m=(2(n-1) + m)/21\}$ for the unweighted and weighted sum-rate cases, respectively, and set the channel fading parameter $\sigma_{m}=2$. Moreover, we assume the power unconstrained case where the energy harvesting rate is $\mu_{n}=6,7,8,9,10,11$ units per slot and a power constrained case where the maximum power constraint is $P_n = 10$ and the energy harvesting rate is  $\mu_{n}=1,2,3,4,5,6$ units per slot. We run the simulation $500$ times to obtain the performance for the different algorithm and policies, as shown in Figs. \ref{fg:ew}, \ref{fg:new}, and \ref{fg:pmax} for the power unconstrained case with ${\cal W}_1$, the power unconstrained case with ${\cal W}_2$, and the power constrained case with ${\cal W}_2$, respectively.

As shown in Fig. \ref{fg:ew}, the maximum throughput in NO-BC is the same as that in O-BC under the optimal energy-bandwidth allocation and the greedy energy policy. This is because in both O-BC and NO-BC, the optimized bandwidth allocation requires that each transmitter only transmit to the receiver with the strongest channel in each slot when the weights are equal (e.g., ${\cal W}_1$), as stated in Theorem \ref{thm:ew} and Corollary \ref{cl:ew}. For the equal bandwidth policy, O-BC performs worse than NO-BC since the NO-BC makes better use of the allocated bandwidth by optimally treating the interference. When we use the unequal weights ${\cal W}_2$, it is seen in Figs. \ref{fg:new} and \ref{fg:pmax} that we may get better performance by using NO-BC instead of O-BC under all policies. However, for the optimal energy-bandwidth allocation, such improvement is quite marginal. Moreover, when the maximum power is constrained, it is seen in Fig. \ref{fg:pmax} that the gap between the performances of the optimal energy-bandwidth allocation and the greedy energy policy decreases as the energy harvesting rates increases.

\subsection{PF Throughput Maximization}
We next evaluate the PF throughput performance in the network with multiple orthogonal broadcast channels. For comparison, we consider three scheduling strategies, namely, the {\em greedy policy}, the {\em traditional PF policy}, and the {\em approximate PF policy}. For the greedy policy, the transmitter evenly splits the maximum available energy for the transmission to each receiver in each slot, i.e., $p_m^k = B_{n}^k/|{\cal M}_n|$, and the equal bandwidth is also allocated, i.e., $a_{m}^k = 1/M$. For the traditional PF policy, the transmitter tries to use the maximum available energy in each slot and one transmission link is chosen to use the entire bandwidth as follows:
\begin{equation}
\arg \max_{m}\Big\{\log(1+p_m^kH_m^k)/\tilde{R}_{m}^k\Big\} \ ,
\end{equation}
where we denote $\tilde{R}_{m}^k$ as the average sum-rate before slot $k$ \cite{CPFSAG}. For the approximate PF policy, we use the approximate PF weights given in \eqref{eq:appPFweight} and then solve a weighted sum-rate maximization problem.

\begin{figure}
\centering
\begin{minipage}[t]{0.49\textwidth}
\includegraphics[width=1\textwidth]{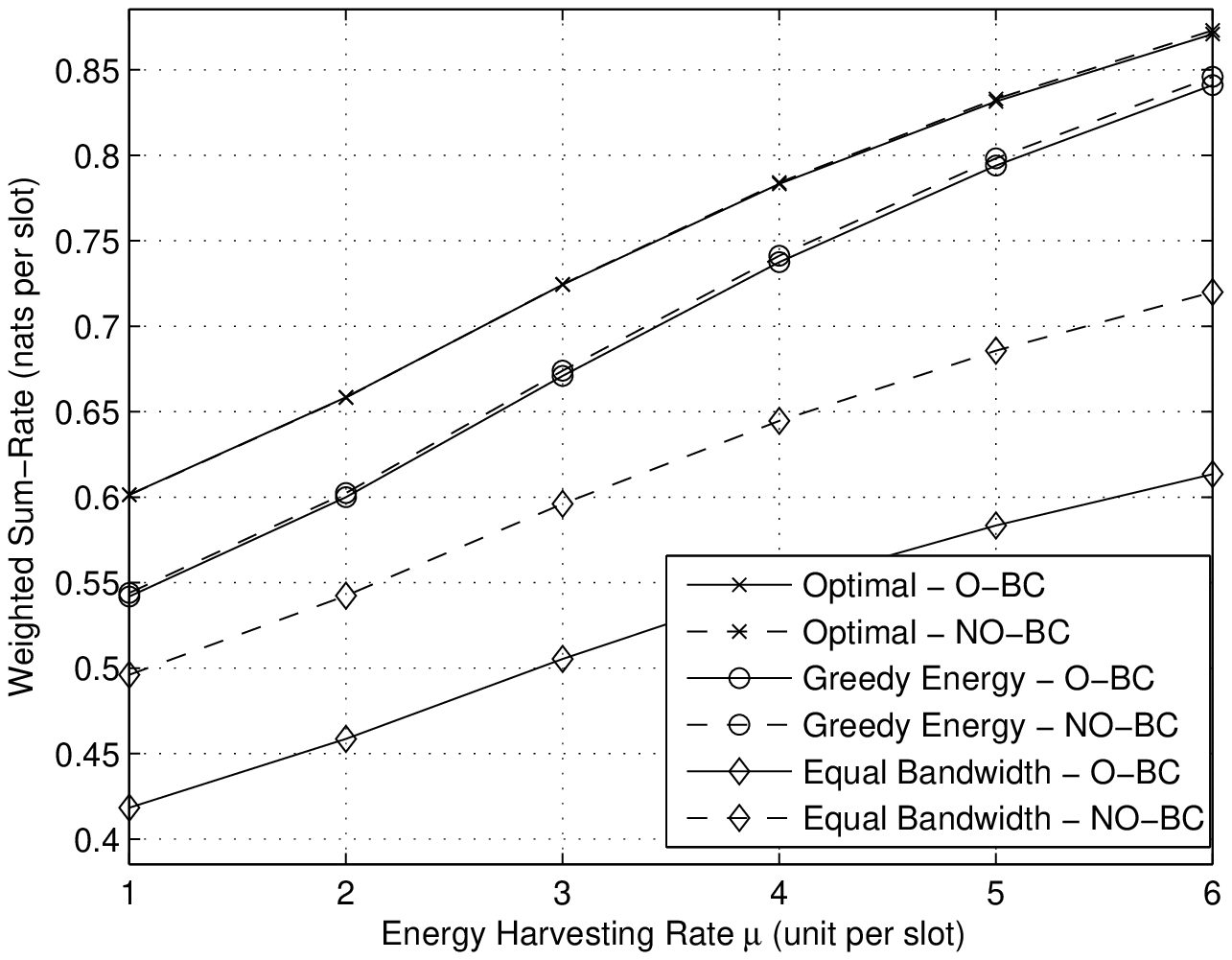}
\caption{Weighted sum-rate comparisons for different policies with the maximum power  (${\cal W}_2$, $P_n=10$).}
\label{fg:pmax}
\end{minipage}
\begin{minipage}[t]{0.49\textwidth}
\includegraphics[width=1\textwidth]{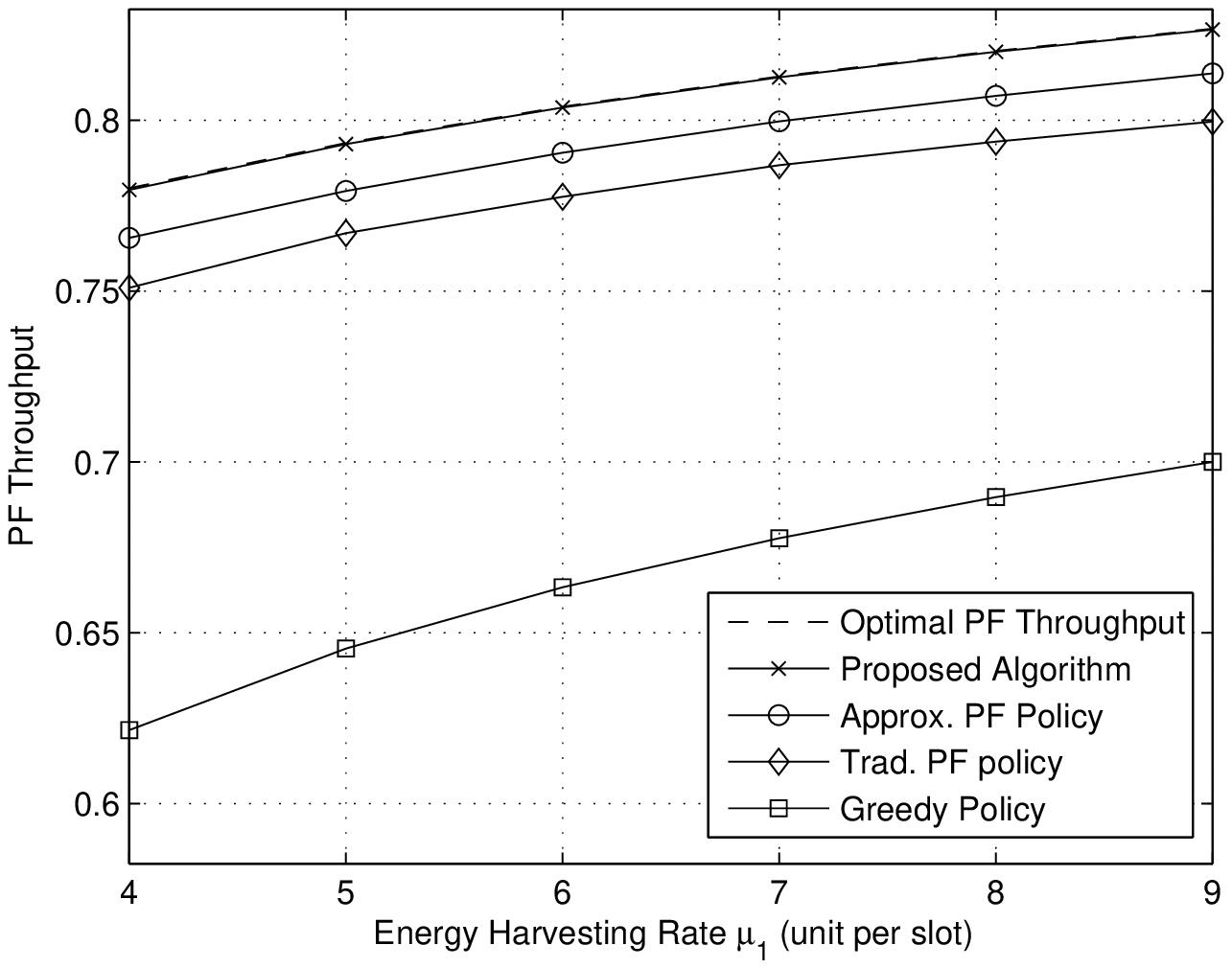}
\caption{Performance comparisons in the varying EH scenario.}
\label{fg:veh}
\end{minipage}

\end{figure}

\begin{figure}
\centering
\begin{minipage}[t]{0.49\textwidth}
\includegraphics[width=1\textwidth]{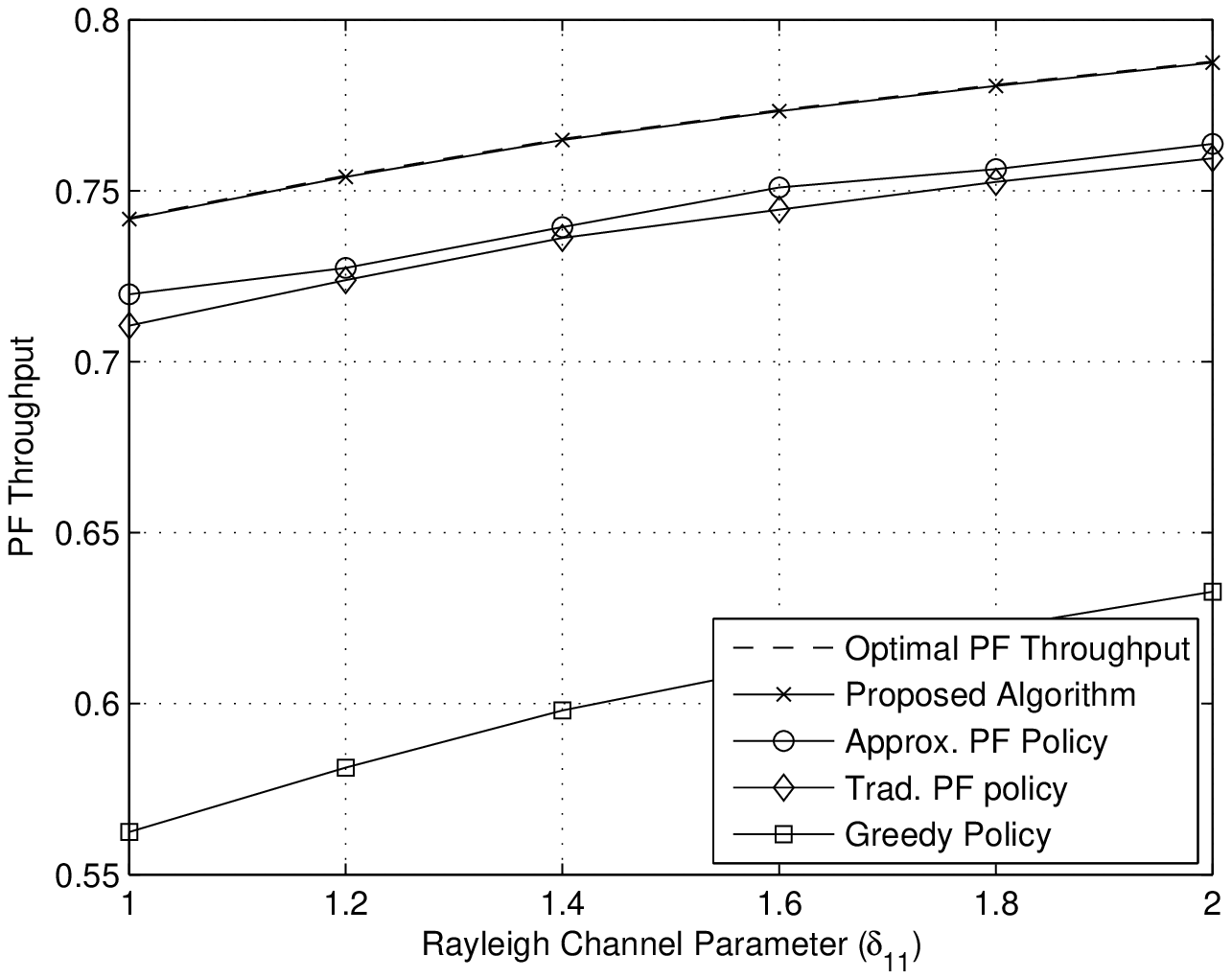}
\caption{Performance comparisons in the varying channel scenario.}
\label{fg:vc}
\end{minipage}
\begin{minipage}[t]{0.49\textwidth}
\includegraphics[width=1\textwidth]{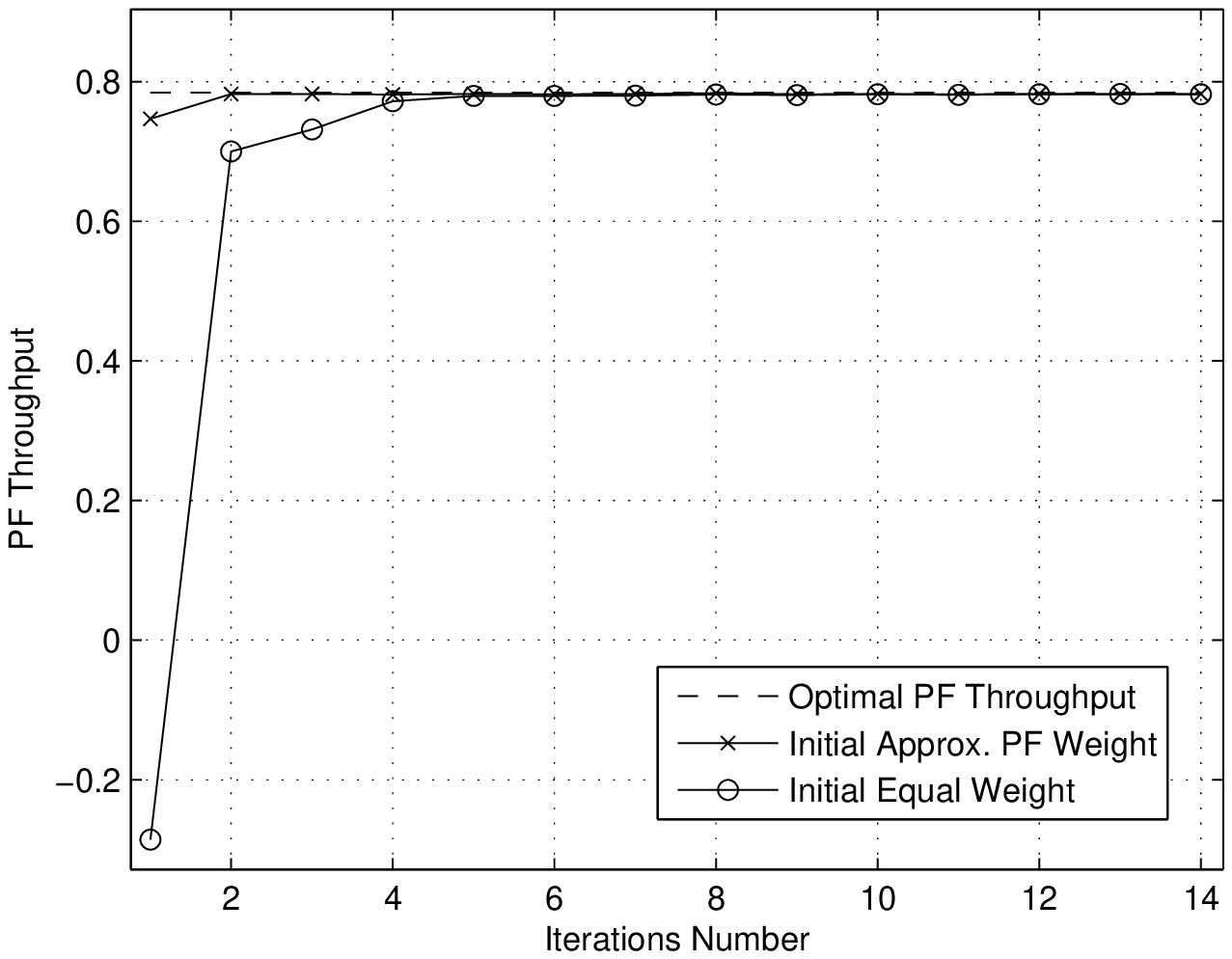}
\caption{Convergence behavior of Algorithm 3.}
\label{fg:conv}
\end{minipage}
\end{figure}

To evaluate the performance of the different algorithm and policies, we consider two scenarios, namely, the {\em varying EH scenario}, where the different transmitters have different means of the energy harvesting such that $\mu_1 + 2 = \mu_2 + 1= \mu_3$ and the channel fading parameter is $\sigma=2$ for all transmitters, and {\em varying channel scenario}, where the different transmitters have different channel fading parameters such that $\sigma_{1(\cdot)} + 0.5 = \sigma_{2(\cdot)}$ and the mean of the energy harvesting is $\mu=2$ for all transmitters. In both the scenarios, the maximum power is unconstrained and we compare the performance of Algorithm 3 and the other three polices with the optimal PF throughput obtained using the generic convex solver. Specifically, in the varying EH scenario and the  varying channel scenario, we assume $\mu_1 = 1,2,3,4,5,6$ units per slot and $\sigma_{1(\cdot)} = 1,1.2,1.4,1.6,1.8,2$, respectively. We run the simulation $500$ times to obtain the performance for the different algorithm and policies, as well as the optimal schedule solved by a general convex solver, as shown in Fig. \ref{fg:veh} and Fig. \ref{fg:vc} for the varying EH scenario and the varying channel scenario, respectively.

%
%

From Fig. \ref{fg:veh} and Fig. \ref{fg:vc}, it is seen that for both scenarios Algorithm 3 achieves the same performance as that achieved by the optimal energy-bandwidth allocation solved by the generic convex solver, which is better than the other policies, as excepted. Specifically, the performance of the approximate PF policy is close to the optimal performance and better than that of the traditional PF and greedy policies. It is because the energy harvesting and channel fading processes are stationary and erodic and the sum-rate achieved by the optimal energy-bandwidth allocation is close to the PF weights parameter. Also, the traditional PF policy is optimal for the transmitters without using the renewable energy source. However, due to the energy harvesting process with the finite battery capacity, the potential energy overflow necessitates the bandwidth share to maximize the proportionally-fair throughput. Therefore, the traditional PF policy gives the suboptimal performance for the transmitters powered by the renewable energy source. Moreover, the greedy policy, which does not take the energy and the fairness factors into account, provides the worst performance among the simulated algorithm/polices.

We also evaluate the convergence speed of Algorithm 3 with different initial weights ${\cal W}$, i.e., the approximate PF weights and equal weights, as shown in Fig. \ref{fg:conv} for $K=20$. It is seen that, the convergence speed with the initial approximate PF weights is faster than that with the initial equal weights, approaching to the optimal performance after around $10$ iterations.

\section{Conclusions}
We have treated the energy-bandwidth allocation problem for a network consisting of multiple energy harvesting transmitters, each broadcasting to multiple receivers, to maximize the weighted throughput and the proportionally fair throughput. Based on the general iterative algorithm developed in \cite{OURFDPAPER} that alternatively solves the energy and bandwidth allocation subproblems, we have developed optimal algorithms for solving the two subproblems for both orthogonal and non-orthogonal broadcast. Moreover, for orthogonal broadcast, we have shown that the PF throughput maximization problem can be converted to the  weighted throughput maximization problem with proper weights. Simulation results demonstrate that the proposed algorithms offer significant performance improvement over various suboptimal allocation schemes. Moreover, it is seen that with energy-harvesting transmitters, non-orthogonal broadcast offers limited gain over orthogonal broadcast.

\appendices
\section{Proof of Proposition \ref{pp:derivative}}

By Lemma \ref{lm:splitting}, we have
\begin{align}
{(F_n^k)}'(p) &=\partial\left(\sum_{m\in{\cal M}_n}W_m\log(1+\frac{p_m^k(p)H_m^k}{H_m^k\sum_{m_0\;|\;H_m^k < H_{m_0}^k}p_{m_0}^k(p)+1})\right)/\partial p\\
&=\partial \left( W_{a}\log(1+\frac{(p - L_a^k) H_{a}^k}{L_a^k H_{a}^k+1})\right)/\partial p\label{eq:nodv}\\
&=\frac{W_{a}}{p + 1/H_{a}^k}\ ,\label{eq:piecewise}\  p\in [L_a^k,L_b^k]
\end{align}
where $\eqref{eq:nodv}$ follows because \eqref{eq:splitting} indicates that, for any $a\in{\cal M}_n$, $p_{a}^k(p)$ is constant when $p<L_a^k$ or $p>L_b^k$.

Hence $(F_n^k)'(p)$ is a piecewise function composed by the segments in the form of $f_m^k(p) \triangleq W_m/(p_m^k+1/H_m^k)$. By Lemma \ref{lm:concave}, ${(F_n^k)}'(p) $ is continuous. Thus, for any two adjacent different cutoff lines $L_a^k < L_b^k$, $L_a^k$ is the intersection of the two curves $f^k_{a}(p)=W_{a}/ (p + 1/H_{a}^k)$ and $f^k_{b}(p)=W_{b}/ (p + 1/H_{b}^k)$. 

Denoting the intersection of $f^k_{a}(p)$ and $f^k_{b}(p)$ as $I_{ab}^k$ (i.e., $p=I_{ab}$ such that $f^k_{a}(I_{ab}^k)=f^k_{b}(I_{ab}^k)$), we then have
\begin{equation}
L_a^k = I_{ab}^k \triangleq \frac{H_{b}^kW_{b} - H_{a}^kW_{a}}{H_{b}^kH_{a}^k\left(W_{a}-W_{b}\right)}\ .
\end{equation}
Specifically, for any $a,b\in{\cal M}_n$, $I_{ab}^k$ is unique if it exists. Then, we can write
\begin{align}
F_n^k(\tilde{p}) &= \int_{0}^{\tilde{p}}(F_n^k)'(p)dp\\
&=\max_{\{ I_{ab}^k<I_{bc}^k<\ldots<\tilde{p}\;|\; a,b,c,\ldots \in{\cal M}_n\}} \left\{\sum_{ab}\int_{I_{ab}^k}^{\min\{I_{bc}^k,\tilde{p}\}}f_a^k(p)dp\right\}\label{eq:optdev}\ ,
\end{align}
where \eqref{eq:optdev} follows since $(F_n^k)'(p)$ is a  piecewise function with the segments of $f_m^k(p)$ and $I_{ab}^k$ is the intersection of $f_a^k(p)$ and $f_b^k(p)$.

\begin{figure}
\centering
\includegraphics[width=.5\textwidth]{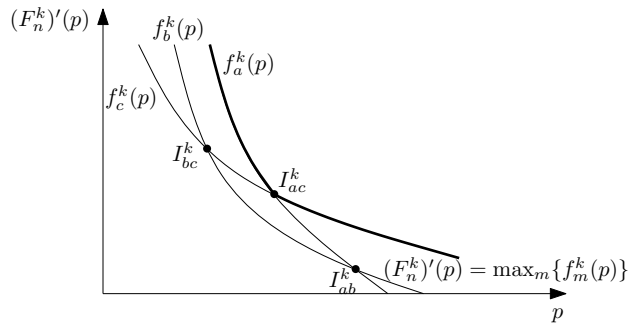}
\caption{The derivative of $F_n^k(p)$.}
\label{fg:optdev}
\end{figure}

Then, as shown in Fig. \ref{fg:optdev}, we can obtain a set of $I_{ab}^k$ and it is easy to verify that the optimal solution to the problem in \eqref{eq:optdev} forms the derivative of $F_n^k(p)$ as
\begin{equation}
{(F_n^k)}'(p)  = \max_{m\in{\cal M}_n}\left\{\frac{W_m}{p + 1/H_m^k}\right\}\ .
\end{equation}

\bibliographystyle{IEEETran}
\bibliography{IEEEabrv,bib}
\end{document}